%% file: main.tex
\documentclass[sigconf]{acmlike}

\usepackage{booktabs} 

\usepackage[all]{xy}

\usepackage{pkg/lhs}
\usepackage{pkg/plots}

\setcopyright{rightsretained}

\copyrightyear{2018}

\begin{document}
\title{Typed Linear Algebra for Efficient Analytical Querying}

\author{J.M. Afonso}
\affiliation{%
  \institution{HASLab/INESC TEC\\University of Minho}
  \city{Braga}
  \country{Portugal}
}
\email{joao.m.afonso@inesctec.pt}

\author{G. Fernandes}
\affiliation{%
  \institution{University of Minho}
  \city{Braga}
  \country{Portugal}
}
\email{a71492@alunos.uminho.pt}

\author{J.P. Fernandes}
\affiliation{%
  \institution{University of Minho}
  \city{Braga}
  \country{Portugal}
}
\email{a70880@alunos.uminho.pt}

\author{F. Oliveira}
\affiliation{%
  \institution{University of Minho}
  \city{Braga}
  \country{Portugal}
}
\email{a57816@alunos.uminho.pt}

\author{B.M. Ribeiro}
\affiliation{%
  \institution{University of Minho}
  \city{Braga}
  \country{Portugal}
}
\email{a71492@alunos.uminho.pt} 
 
\author{R. Pontes}
\affiliation{%
  \institution{HASLab/INESC TEC\\University of Minho}
  \city{Braga}
  \country{Portugal}
}
\email{rapontes@inesctec.pt}

\author{J.N. Oliveira}
\affiliation{%
  \institution{HASLab/INESC TEC\\University of Minho}
  \city{Braga}
  \country{Portugal}
}
\email{jno@di.uminho.pt}

\author{A.J. Proen\c{c}a}
\affiliation{%
  \institution{Algoritmi Center\\University of Minho}
  \city{Braga}
  \country{Portugal}
}
\email{aproenca@di.uminho.pt}

\renewcommand{\shortauthors}{Afonso et al.}

\begin{abstract}
This paper uses typed linear algebra (LA) to represent data
and perform analytical querying in a single, unified framework. The typed
approach offers strong type checking (as in modern programming languages)
and a diagrammatic way of expressing queries (paths in LA diagrams). A kernel
of LA operators has been implemented so that paths extracted from LA diagrams
can be executed. The approach is validated and evaluated
taking TPC-H benchmark queries as reference.
The performance of the LA-based approach is compared with popular database competitors (PostgreSQL and MySQL).
\end{abstract}

%
%



\maketitle

\input{sections/1-intro}
\input{sections/2-encoding}
\input{sections/3-la-encoding}
\input{sections/4-incremental}
\input{sections/5-SQL-LA}
\input{sections/6-implementation}

\input{sections/7-benchmarks}

\input{sections/8-conclusions}

\begin{acks}
This work was supported by the EU Framework Programme for Research and Innovation 2014-2020, Horizon 2020, under grant agreement No. 732051 (\textsf{CloudDB Appliance} project) and by Project Search-ON2 (NORTE-07-0162- FEDER-000086), co-funded by the North Portugal Regional Operational Programme (ON.2 - O Novo Norte), under the National Strategic Reference Framework, through the European Regional Development Fund. 
\end{acks}

\bibliographystyle{ACM-Reference-Format}
\bibliography{sample-bibliography}

\input{sections/appendix}

\end{document}

%% file: sections/1-intro.tex
\section{Introduction}
{\vskip 1em\hfill
\begin{minipage}{0.41\textwidth}\em\small
``Only by taking infinitesimally small units for observation (the \emph{differential} of history, that is, the individual tendencies of men) and attaining to the art of \emph{integrating} them (that is, finding the sum of these infinitesimals) can we hope to arrive at the laws of history."
\vskip 1ex
\em \flushright \small Leo Tolstoy, ``War and Peace"%
\flushright \vskip -0.8ex  - Book XI, Chap.II (1869)
\end{minipage}
}

\vskip 1em
J. Gantz \emph{et al} predict that the amount of digital bits created and consumed each year in the USA will grow to 6.6 zettabytes by 2020 \cite{GR12}. 
In this age of the information society, \emph{data records} are the \emph{differentials} of Tolstoy's quote given above.
Through analytical querying, strategy planners are mining such infinitesimals
and \emph{integrating} them so as to infer the ``laws of our society". A century
and a half in advance, Tolstoy has given us perhaps the oldest definition
of data mining that we have.


Since the early days of psychometrics in the {social sciences} (1970s),
\emph{linear algebra} (LA) has been central to data analysis, namely through
tensor decomposition, incremental tensor analysis \cite{STPYF08} and so on. Abadir and Magnus
\cite{AM05} stress the need for a \emph{standardized} notation for linear algebra
in the field of econometrics and statistics.

More recently, it has been shown that data consolidation can be elegantly
expressed in \emph{typed} linear algebra \cite{MO15,OM17}, a categorial approach
to linear algebra \cite{MO13c} in which matrices are represented by arrows
\ensuremath{\rarrow{\Varid{m}}{\Conid{M}}{\Varid{n}}} where \ensuremath{\Conid{M}} is a LA-expression denoting a matrix, \ensuremath{\Varid{m}} is the
number of columns of \ensuremath{\Conid{M}} and \ensuremath{\Varid{n}} is the corresponding number of rows. The
approach is \emph{strongly typed} in the sense that it is free from matrix
dimension errors, by construction \cite{MO13c}. Moreover, it is \emph{type-polymorphic},
making room for proving properties of data constructions relying on types alone.
For instance, the ``free theorem" \cite{Wa89} of the \emph{data cube} operator
given in \cite{OM17} is proved in that way.

The main aim of \cite{MO15,OM17} was to show how analytical querying theory
benefits from the typed LA approach, essentially from a foundation point
of view. In the current paper, we propose, validate and evaluate such a typed LA approach as a means of data analysis programming.

We implemented a minimal kernel of LA operators needed
when scripting analytic queries. Such scripts essentially describe paths of diagrams
whose arrows record data encoded as typed matrices. We also show how to infer such LA scripts from standard SQL code. Finally, we evaluate the approach
by running the scripts of some queries of the TPC-H benchmark suite.
This gives evidence of such scripts being faster and more efficient than standard SQL-based solutions.

Our main contributions are:
\begin{itemize}
    \item A \emph{typed linear algebra} approach 
to complex data querying based on a minimal LA kernel.
    \item A new way to build query plans as \emph{paths} of typed LA diagrams, ensuring \emph{type-correctness} by construction.
    \item An evaluation of the approach using queries of the TPC-H benchmark suite, which includes a comparison with two widely used and industry proven databases, PostgreSQL\cite{PostgreSQL_doc} and MySQL \cite{MySQL_doc}. Although our system is still a prototype, its measured performance is better in the majority of the benchmarks.
\end{itemize}

%% file: sections/2-encoding.tex
\section{Algebraic encoding of data}
As starting point for describing our approach, consider the following samples of two relational data
tables randomly generated by the TPC-H benchmark suite \cite{TPC17} --- table
\ensuremath{\mathsf{\Varid{orders}}}
\begin{eqnarray*}
\begin{minipage}{\textwidth}\fontfamily{ptm}\selectfont{\fontsize{1.90ex}{1.90ex}\selectfont 
\begin{tabbing}\ttfamily
~\char35{}~\char124{}~o\char95{}orderkey~\char124{}~o\char95{}orderpriority~\char124{}~o\char95{}orderdate\\
\ttfamily ~\char45{}\char45{}\char43{}\char45{}\char45{}\char45{}\char45{}\char45{}\char45{}\char45{}\char45{}\char45{}\char45{}\char45{}\char45{}\char43{}\char45{}\char45{}\char45{}\char45{}\char45{}\char45{}\char45{}\char45{}\char45{}\char45{}\char45{}\char45{}\char45{}\char45{}\char45{}\char45{}\char45{}\char43{}\char45{}\char45{}\char45{}\char45{}\char45{}\char45{}\char45{}\char45{}\char45{}\char45{}\char45{}\char45{}\\
\ttfamily ~1~\char124{}~~~~~~~5699~\char124{}~~~~~~~~~~2\char45{}HIGH~\char124{}~~1992\char45{}07\char45{}30\\
\ttfamily ~2~\char124{}~~~~~~~4354~\char124{}~~~~~~~~3\char45{}MEDIUM~\char124{}~~1994\char45{}09\char45{}30\\
\ttfamily ~3~\char124{}~~~~~~~~551~\char124{}~~~~~~~~~~2\char45{}HIGH~\char124{}~~1995\char45{}05\char45{}30\\
\ttfamily ~4~\char124{}~~~~~~~2723~\char124{}~~~~~~~~~~2\char45{}HIGH~\char124{}~~1995\char45{}10\char45{}06\\
\ttfamily ~5~\char124{}~~~~~~~3392~\char124{}~~~~~~~~3\char45{}MEDIUM~\char124{}~~1995\char45{}10\char45{}28
\end{tabbing}
}
\end{minipage}
\end{eqnarray*}
and table \ensuremath{\mathsf{\Varid{lineitem}}}:
\begin{eqnarray*}
\begin{minipage}{\textwidth}\fontfamily{ptm}\selectfont{\fontsize{1.79ex}{1.79ex}\selectfont 
\begin{tabbing}\ttfamily
~~~\char35{}~\char124{}~l\char95{}orderkey~\char124{}~l\char95{}quantity~\char124{}~l\char95{}linestatus~\char124{}~l\char95{}extendedprice\\
\ttfamily ~~~\char45{}\char45{}\char43{}\char45{}\char45{}\char45{}\char45{}\char45{}\char45{}\char45{}\char45{}\char45{}\char45{}\char45{}\char45{}\char43{}\char45{}\char45{}\char45{}\char45{}\char45{}\char45{}\char45{}\char45{}\char45{}\char45{}\char45{}\char45{}\char43{}\char45{}\char45{}\char45{}\char45{}\char45{}\char45{}\char45{}\char45{}\char45{}\char45{}\char45{}\char45{}\char45{}\char45{}\char43{}\char45{}\char45{}\char45{}\char45{}\char45{}\char45{}\char45{}\char45{}\char45{}\char45{}\char45{}\char45{}\char45{}\char45{}\char45{}\char45{}\\
\ttfamily ~~~1~\char124{}~~~~~~~2723~\char124{}~~~~~~~4\char46{}00~\char124{}~O~~~~~~~~~~~~\char124{}~~~~~~~~~2124\char46{}32\\
\ttfamily ~~~2~\char124{}~~~~~~~~551~\char124{}~~~~~~~1\char46{}00~\char124{}~O~~~~~~~~~~~~\char124{}~~~~~~~~16994\char46{}56\\
\ttfamily ~~~3~\char124{}~~~~~~~5699~\char124{}~~~~~~~2\char46{}00~\char124{}~F~~~~~~~~~~~~\char124{}~~~~~~~~32735\char46{}70\\
\ttfamily ~~~4~\char124{}~~~~~~~4354~\char124{}~~~~~~~9\char46{}00~\char124{}~O~~~~~~~~~~~~\char124{}~~~~~~~~~1902\char46{}10\\
\ttfamily ~~~5~\char124{}~~~~~~~3392~\char124{}~~~~~~~1\char46{}00~\char124{}~F~~~~~~~~~~~~\char124{}~~~~~~~~42846\char46{}80\\
\ttfamily ~~~6~\char124{}~~~~~~~4354~\char124{}~~~~~~~5\char46{}00~\char124{}~F~~~~~~~~~~~~\char124{}~~~~~~~~35707\char46{}22\\
\ttfamily ~~~7~\char124{}~~~~~~~5699~\char124{}~~~~~~~5\char46{}00~\char124{}~F~~~~~~~~~~~~\char124{}~~~~~~~~44064\char46{}48\\
\ttfamily ~~~8~\char124{}~~~~~~~3392~\char124{}~~~~~~~3\char46{}00~\char124{}~O~~~~~~~~~~~~\char124{}~~~~~~~~~7168\char46{}84
\end{tabbing}
}
\end{minipage}
\end{eqnarray*}
Further consider
the following, much simplified
version of query number 3 of the same benchmark:
\begin{hscode}\SaveRestoreHook
\column{B}{@{}>{\hspre}l<{\hspost}@{}}%
\column{4}{@{}>{\hspre}l<{\hspost}@{}}%
\column{E}{@{}>{\hspre}l<{\hspost}@{}}%
\>[B]{} \sql{select} \;{}\<[E]%
\\
\>[B]{}\hsindent{4}{}\<[4]%
\>[4]{}\Varid{o\char95 orderpriority},{}\<[E]%
\\
\>[B]{}\hsindent{4}{}\<[4]%
\>[4]{}\Varid{o\char95 orderdate},{}\<[E]%
\\
\>[B]{}\hsindent{4}{}\<[4]%
\>[4]{}\Varid{sum}\;(\Varid{l\char95 quantity}\mathbin{*}\Varid{l\char95 extendedprice}){}\<[E]%
\\
\>[B]{} \sql{from} \;{}\<[E]%
\\
\>[B]{}\hsindent{4}{}\<[4]%
\>[4]{}\Varid{lineitem},{}\<[E]%
\\
\>[B]{}\hsindent{4}{}\<[4]%
\>[4]{}\Varid{orders}{}\<[E]%
\\
\>[B]{}\mathbf{where}{}\<[E]%
\\
\>[B]{}\hsindent{4}{}\<[4]%
\>[4]{}\Varid{l\char95 orderkey}\mathrel{=}\Varid{o\char95 orderkey}{}\<[E]%
\\
\>[B]{} \sql{group\ by} \;{}\<[E]%
\\
\>[B]{}\hsindent{4}{}\<[4]%
\>[4]{}\Varid{o\char95 orderdate},{}\<[E]%
\\
\>[B]{}\hsindent{4}{}\<[4]%
\>[4]{}\Varid{o\char95 orderpriority}{}\<[E]%
\ColumnHook
\end{hscode}\resethooks

In general, given a table \ensuremath{\mathsf{\Varid{t}}}, one of its attributes \ensuremath{\Varid{a}} and the number \ensuremath{\Varid{i}} of one of its records (rows),
we denote the corresponding data value by the expression \ensuremath{\lkp{\Varid{t}}{\Varid{a}}{\Varid{i}}}. For instance,
\ensuremath{\lkp{\Varid{orders}}{\Varid{o\char95 orderkey}}{\mathrm{3}}\mathrel{=}\mathrm{551}} above.

Our typed LA encoding of data is inherently \emph{columnar} \cite{FPST11}.
As is usual, we shall split columns (attributes) into two groups: the so-called
\emph{dimensions} and the so-called \emph{measures}.

Every \emph{dimension} attribute \ensuremath{\Varid{d}} of a given table \ensuremath{\mathsf{\Varid{t}}} is represented by
a so-called \emph{projection} function \ensuremath{\proj{\Varid{t}}{\Varid{d}}\mathbin{:}\card{\Varid{t}}\to |\Varid{d}|},
where \ensuremath{|\Varid{d}|} is the type containing the range of values of attribute \ensuremath{\Varid{d}},
and \ensuremath{\card{\Varid{t}}\mathrel{=}\{\mskip1.5mu \mathrm{1},\mathbin{...},\card{\Varid{t}}\mskip1.5mu\}} is the set of row indices of \ensuremath{\Varid{t}}.\footnote{
Note the slight abuse of notation, which also enables
us to define the singleton type by \ensuremath{\mathrm{1}\mathrel{=}\{\mskip1.5mu \mathrm{1}\mskip1.5mu\}}, useful in the sequel.
As expected, \ensuremath{\mathrm{0}\mathrel{=}\{\mskip1.5mu \mskip1.5mu\}}.}
Clearly,
\begin{eqnarray}
	\ensuremath{\proj{\Varid{t}}{\Varid{d}}\;(\Varid{i})\mathrel{=}\lkp{\Varid{t}}{\Varid{d}}{\Varid{i}}}
	\label{eq:180228b}
\end{eqnarray}
e.g. \ensuremath{\proj{\Varid{orders}}{\Varid{orderkey}}\;(\mathrm{3})\mathrel{=}\lkp{\Varid{orders}}{\Varid{orderkey}}{\mathrm{3}}\mathrel{=}\mathrm{551}}.\footnote{Since notation \ensuremath{\proj{\Varid{t}}{\Varid{d}}} makes it clear which table an attribute belongs to, we henceforth drop TPC-H prefixes ``\text{\ttfamily o\char95{}}`" and ``\text{\ttfamily l\char95{}}" f from attribute names.}

Projection functions represent relational data sets with no loss of information
insofar as the original tuples are concerned, as these can be recovered from them by a function combinator
called \emph{pairing}. Let \ensuremath{\Varid{f}\mathbin{:}\Conid{A}\to \Conid{B}} and \ensuremath{\Varid{g}\mathbin{:}\Conid{A}\to \Conid{C}} be functions with
the same source type. We denote by \ensuremath{{\Varid{f}}\kr{\Varid{g}}\mathbin{:}\Conid{A}\to \Conid{B} \times \Conid{C}} the \emph{pairing}
of \ensuremath{\Varid{f}} and \ensuremath{\Varid{g}}, that is, the function defined by
\begin{eqnarray}
	\ensuremath{({\Varid{f}}\kr{\Varid{g}})\;\Varid{a}\mathrel{=}(\Varid{f}\;\Varid{a},\Varid{g}\;\Varid{a})}
	\label{eq:180214b}
\end{eqnarray}
For instance, abbreviating \ensuremath{\mathsf{\Varid{orders}}} to \ensuremath{\Varid{o}} for economy of notation:
\begin{quote}
	\ensuremath{({\proj{\Varid{o}}{\Varid{orderpriority}}}\kr{\proj{\Varid{o}}{\Varid{orderdate}}})\;\mathrm{3}} =
\\\rule{12em}{0pt}
		\ensuremath{(\text{\ttfamily '2-HIGH'},\text{\ttfamily '1995-05-30'})},
\\
	\ensuremath{({\proj{\Varid{o}}{\Varid{orderkey}}}\kr{({\proj{\Varid{o}}{\Varid{orderpriority}}}\kr{\proj{\Varid{o}}{\Varid{orderdate}}})})\;\mathrm{3}\mathrel{=}}
\\\rule{9em}{0pt}
	\ensuremath{(\mathrm{551},(\text{\ttfamily '2-HIGH'},\text{\ttfamily '1995-05-30'}))},
\end{quote}
and so on.

Since columns (dimension attributes) ``are" functions, we can represent the
schema of the (simplified) TPC-H data base by a diagram of function types, where
we associate type variables to the ranges of the corresponding attributes:
\begin{eqnarray}
\myxym{
&
	\ensuremath{\R}
\\
	\ensuremath{\Conid{S}}
&
	\ensuremath{\card{\Varid{lineitem}}}
		\ar[l]^{\ensuremath{\proj{\Varid{l}}{\Varid{linestatus}}}}
		\ar[u]^{\ensuremath{\proj{\Varid{l}}{\Varid{extendedprice}}}}
		\ar[d]^{\ensuremath{\proj{\Varid{l}}{\Varid{orderkey}}}}
		\ar[r]^{\ensuremath{\proj{\Varid{l}}{\Varid{quantity}}}}
&
	\ensuremath{\R}
\\
&
	\ensuremath{\Conid{K}}
&
	\ensuremath{\card{\Varid{orders}}}
	\ar[rr]^{\ensuremath{\proj{\Varid{o}}{\Varid{orderpriority}}}}
	\ar[d]^{\ensuremath{\proj{\Varid{o}}{\Varid{orderdate}}}}
	\ar[l]_{\ensuremath{\proj{\Varid{o}}{\Varid{orderkey}}}}
&&
	\ensuremath{\Conid{P}}
\\
&
&
	\ensuremath{\Conid{D}}
&
}
\end{eqnarray}
For instance,
\ensuremath{|\Varid{l\char95 orderkey}|\mathrel{=}\Conid{K}} and so \ensuremath{\proj{\Varid{l}}{\Varid{orderkey}}\mathbin{:}\card{\Varid{l}}\to \Conid{K}}.
That the target type of \ensuremath{\proj{\Varid{l}}{\Varid{orderkey}}} has to be the same as that
of \ensuremath{\proj{\Varid{o}}{\Varid{orderkey}}}, i.e.\ \ensuremath{\Conid{K}}, is entailed by the clause
\ensuremath{\Varid{l\char95 orderkey}\mathrel{=}\Varid{o\char95 orderkey}} in the query, otherwise this clause wouldn't type.

Besides \ensuremath{\Conid{K}} (for Key), the other type variables associated to each of the
SQL concrete types present in the model \cite{TPC17} have the following meaning:
\ensuremath{\Conid{P}} stands for \ensuremath{\Conid{Priority}},
\ensuremath{\Conid{D}} for Date,
and \ensuremath{\Conid{S}} for Status.
\ensuremath{\R} denotes the set of all real numbers, where prices are valuated.

One striking observation about the diagram above is that all types are ranges of values
(and therefore finite) with the exception of \ensuremath{\R}, the target type of \ensuremath{\proj{\Varid{l}}{\Varid{extendedprice}}}
and \ensuremath{\proj{\Varid{l}}{\Varid{quantity}}}, corresponding to DECIMAL types in the SQL code. Attributes of this kind should not be regarded
as \emph{dimensions} but rather as \emph{measures} because one can operate over them to
produce consolidated data. The minimum algebraic structure for consolidation to take
place is that of a \emph{semiring}, offering a multiplicative (\ensuremath{ \times }) and an additive operator (+)
with the expected properties (as in \ensuremath{\R}), notably \ensuremath{\Varid{a} \times (\Varid{b}\mathbin{+}\Varid{c})\mathrel{=}\Varid{a} \times \Varid{b}\mathbin{+}\Varid{a} \times \Varid{c}} ensuring \emph{linearity}.

Treating measures in the same way as dimensions would have the disadvantage that
data consolidation has to be carried and reasoned out explicitly, involving
lots of (nested) summations and quantifiers alike. To circumvent this problem,
in the following section we generalize projection functions to matrices and
move from the (functional) \ensuremath{\lambda }-calculus to the realm of matrix calculi,
i.e., linear algebra.

%% file: sections/3-la-encoding.tex
\section{Linear algebraic encoding of data} \label{sec:180219a}
The typed LA approach to data representation \cite{OM17} consists in representing
projection functions by (Boolean) matrices, as follows: let \ensuremath{\Varid{f}\mathbin{:}\Conid{A}\to \Conid{B}} be any
function, where \ensuremath{\Conid{A}} and \ensuremath{\Conid{B}} are finite. Function \ensuremath{\Varid{f}} can be represented
by a matrix \ensuremath{\mean{\Varid{f}}} with \ensuremath{\Conid{A}}-many columns and \ensuremath{\Conid{B}}-many rows such that, for
any \ensuremath{\Varid{b}\;{\in}\;\Conid{B}} and \ensuremath{\Varid{a}\;{\in}\;\Conid{A}}, matrix cell \footnote{Following the infix notation
usually adopted for relations (which are Boolean matrices), for instance
\ensuremath{\Varid{y}\leq \Varid{x}}, we write \ensuremath{\Varid{y}\;\Conid{M}\;\Varid{x}} to denote the contents of the cell in matrix \ensuremath{\Conid{M}}
addressed by row \ensuremath{\Varid{y}} and column \ensuremath{\Varid{x}}.
}
\ensuremath{\cell{\Varid{b}}{\mean{\Varid{f}}}{\Varid{a}}\mathrel{=}\mathrm{1}} if \ensuremath{\Varid{b}\mathrel{=}\Varid{f}\;\Varid{a}}, otherwise \ensuremath{\cell{\Varid{b}}{\mean{\Varid{f}}}{\Varid{a}}\mathrel{=}\mathrm{0}}.
For instance, \ensuremath{\mean{\proj{\Varid{o}}{\Varid{orderpriority}}}} is the matrix
\begin{eqnarray}
\begin{array}{r|ccccc}&{1}&{2}&{3}&{4}&{5}\\\hline \hbox{2-HIGH}&1&0&1&1&0\\\hbox{3-MEDIUM}&0&1&0&0&1
\end{array}
	\label{eq:140403a}
\end{eqnarray}
and \ensuremath{\mean{\proj{\Varid{o}}{\Varid{orderdate}}}} is the matrix:
\begin{eqnarray*}
\begin{array}{r|ccccc}&{1}&{2}&{3}&{4}&{5}\\\hline \hbox{1992-07-30}&1&0&0&0&0\\\hbox{1994-09-30}&0&1&0&0&0\\\hbox{1995-05-30}&0&0&1&0&0\\\hbox{1995-10-06}&0&0&0&1&0\\\hbox{1995-10-28}&0&0&0&0&1
\end{array}
\end{eqnarray*}
Likewise, it can be easily shown that
\begin{eqnarray}
	\ensuremath{\cell{(\Varid{b},\Varid{c})}{\mean{{\Varid{f}}\kr{\Varid{g}}}}{\Varid{a}}} = \ensuremath{(\cell{\Varid{b}}{\mean{\Varid{f}}}{\Varid{a}}) \times (\cell{\Varid{c}}{\mean{\Varid{g}}}{\Varid{a}})}
\end{eqnarray}
holds, since multiplication within \ensuremath{\{\mskip1.5mu \mathrm{0},\mathrm{1}\mskip1.5mu\}} implements logic conjunction.

As in \cite{OM17} we shall abuse of notation and (very conveniently, as we
shall see) drop the parentheses from \ensuremath{\mean{\Varid{f}}}. This is consistent with writing
\ensuremath{{\Varid{f}}\kr{\Varid{g}}} to denote the operation above, which in fact corresponds to a well-known
matrix operator. It is called the \emph{Khatri-Rao product} \cite{RR:98}
\ensuremath{{\Conid{M}}\kr{\Conid{N}}} of two arbitrary matrices \ensuremath{\Conid{M}} and \ensuremath{\Conid{N}} and is defined index-wise by:
\begin{eqnarray}
	\ensuremath{(\Varid{b},\Varid{c})\;({\Conid{M}}\kr{\Conid{N}})\;\Varid{a}} &=& \ensuremath{(\Varid{b}\;\Conid{M}\;\Varid{a})} \times \ensuremath{(\Varid{c}\;\Conid{N}\;\Varid{a})}
	\label{eq:130129b}
\end{eqnarray}
Thus \ensuremath{{\mean{\proj{\Varid{o}}{\Varid{orderdate}}}}\kr{\mean{\proj{\Varid{o}}{\Varid{orderpriority}}}}} is the typed matrix:
\small
\begin{eqnarray*}
\begin{array}{r|ccccc}&{1}&{2}&{3}&{4}&{5}\\\hline \hbox{(1992-07-30,2-HIGH)}&1&0&0&0&0\\\hbox{(1992-07-30,3-MEDIUM)}&0&0&0&0&0\\\hbox{(1994-09-30,2-HIGH)}&0&0&0&0&0\\\hbox{(1994-09-30,3-MEDIUM)}&0&1&0&0&0\\\hbox{(1995-05-30,2-HIGH)}&0&0&1&0&0\\\hbox{(1995-05-30,3-MEDIUM)}&0&0&0&0&0\\\hbox{(1995-10-06,2-HIGH)}&0&0&0&1&0\\\hbox{(1995-10-06,3-MEDIUM)}&0&0&0&0&0\\\hbox{(1995-10-28,2-HIGH)}&0&0&0&0&0\\\hbox{(1995-10-28,3-MEDIUM)}&0&0&0&0&1
\end{array}
\end{eqnarray*}
\normalsize


Matrices representing projection functions can be chained with each
other (thus yielding \emph{queries}, as we shall see) thanks to two main LA operations:
\emph{composition} and \emph{converse}. Given two functions \ensuremath{\Varid{g}\mathbin{:}\Conid{A}\to \Conid{B}} and \ensuremath{\Varid{f}\mathbin{:}\Conid{B}\to \Conid{C}},
their \emph{composition} \ensuremath{\Varid{f} \comp \Varid{g}} is defined by
\begin{quote}
	\ensuremath{(\Varid{f} \comp \Varid{g})\;\Varid{a}\mathrel{=}\Varid{f}\;(\Varid{g}\;\Varid{a})}.
\end{quote}
Matrix-wise, one can define \ensuremath{\mean{\Varid{g}} \comp \mean{\Varid{f}}} too, where we overload the dot to also mean
matrix \emph{multiplication}. In general, given two matrices \ensuremath{\Conid{N}\mathbin{:}\Conid{A}\to \Conid{B}}
and \ensuremath{\Conid{M}\mathbin{:}\Conid{B}\to \Conid{C}}, their \emph{composition} (or matrix-matrix multiplication, \textsc{mmm}) is the matrix
\ensuremath{\Conid{M} \comp \Conid{N}} defined by:
\begin{eqnarray}
	\ensuremath{\cell{\Varid{c}}{(\Conid{M} \comp \Conid{N})}{\Varid{a}}} &=& \ensuremath{\rcb{\Sigma }{\Varid{b}}{}{\cell{\Varid{c}}{\Conid{M}}{\Varid{b}} \times \cell{\Varid{b}}{\Conid{N}}{\Varid{a}}}}
	\label{eq:120427a}
\end{eqnarray}

Note how we extend the arrow notation used to type functions to also
type arbitrary matrices, \ensuremath{\Conid{M}\mathbin{:}\Conid{A}\to \Conid{B}} meaning that matrix \ensuremath{\Conid{A}} has \ensuremath{\Conid{A}}-many
columns and \ensuremath{\Conid{B}}-many rows.
Writing $\larrow A M B$ means the same as $\rarrow A M B$ or as \ensuremath{\Conid{M}\mathbin{:}\Conid{A}\to \Conid{B}}.

Wherever a matrix has one sole row it is said
to be a \emph{row vector} and we write e.g.\ \ensuremath{\Varid{v}\mathbin{:}\Conid{A}\to \mathrm{1}} to say this. Type
\ensuremath{\mathrm{1}\mathrel{=}\{\mskip1.5mu \mathrm{1}\mskip1.5mu\}} is the singleton type whose unique element is number \ensuremath{\mathrm{1}}, as we have already said.
Clearly, type \ensuremath{\mathrm{1}} is monomorphic.

Given a type \ensuremath{\Conid{A}}, there is a unique row vector wholly filled with \ensuremath{\mathrm{1}}s.
This is termed ``bang'' \cite{OM17} and denoted by \ensuremath{{\bang}\mathbin{:}\Conid{A}\to \mathrm{1}}. Clearly,
\begin{eqnarray}
	\ensuremath{{\Conid{M}}\kr{{\bang}}\mathrel{=}\Conid{M}\mathrel{=}{{\bang}}\kr{\Conid{M}}}
	\label{eq:140602a}
\end{eqnarray}
cf. e.g.
\begin{eqnarray*}
	\ensuremath{{\matrix{\mathrm{2}}{\mathrm{3}}{\mathrm{4}}{\mathrm{5}}}\kr{\begin{bmatrix}\mathrm{1}&\mathrm{1}\end{bmatrix}}\mathrel{=}\matrix{\mathrm{2} \times \mathrm{1}}{\mathrm{3} \times \mathrm{1}}{\mathrm{4} \times \mathrm{1}}{\mathrm{5} \times \mathrm{1}}\mathrel{=}\matrix{\mathrm{2}}{\mathrm{3}}{\mathrm{4}}{\mathrm{5}}}
\end{eqnarray*}

There is also
a unique square matrix of type \ensuremath{\Conid{A}\to \Conid{A}} whose diagonal is wholly filled with
\ensuremath{\mathrm{1}}s and otherwise filled with \ensuremath{\mathrm{0}}s --- it is termed the \emph{identity} matrix
and is denoted by \ensuremath{\Varid{id}\mathbin{:}\Conid{A}\to \Conid{A}}. This is the unit of matrix composition:
\begin{quote}
	\ensuremath{\Conid{M} \comp \Varid{id}\mathrel{=}\Conid{M}\mathrel{=}\Varid{id} \comp \Conid{M}}.
\end{quote}

The other capital LA operator for building analytical que\-ries is \emph{transposition},
or \emph{converse} \ensuremath{\conv{\Conid{M}}\mathbin{:}\Conid{B}\to \Conid{A}}, given matrix \ensuremath{\Conid{M}\mathbin{:}\Conid{A}\to \Conid{B}}. It is defined
by swapping rows with columns:
\ensuremath{\cell{\Varid{a}}{\conv{\Conid{M}}}{\Varid{b}}} = \ensuremath{\cell{\Varid{b}}{\Conid{M}}{\Varid{a}}}.
The following laws hold:
\( \conv{(\conv M)} = M \) (idempotence) and
\(
	\conv{(M \comp N)} = \conv N \comp \conv M
\)
(contravariance).
Clearly, the converse of a \emph{row} vector \ensuremath{\Varid{v}\mathbin{:}\Conid{A}\to \mathrm{1}} is a \emph{column}
vector \ensuremath{\conv{\Varid{v}}\mathbin{:}\mathrm{1}\to \Conid{A}}. A row vector which is also a column vector is a
\emph{scalar}. For instance, given \ensuremath{{\bang}\mathbin{:}\Conid{A}\to \mathrm{1}}, the scalar
\ensuremath{\larrow{\mathrm{1}}{{\bang} \comp \conv{{\bang}}}{\mathrm{1}}} counts the number of elements of the finite
type \ensuremath{\Conid{A}}.

Vectors provide a very convenient LA representation of measure attributes,
as we shall see shortly. Before this, we introduce another matrix product of
capital importance --- the so-called Hadamard product:
\begin{eqnarray}
	\ensuremath{\Varid{b}\;(\Conid{M} \times \Conid{N})\;\Varid{a}\mathrel{=}(\Varid{b}\;\Conid{M}\;\Varid{a}) \times (\Varid{b}\;\Conid{N}\;\Varid{a})}
	\label{eq:180222a}
\end{eqnarray}
Typewise, \ensuremath{\Conid{M}}, \ensuremath{\Conid{N}} and \ensuremath{\Conid{M} \times \Conid{N}} are all of the same type, for instance
\begin{eqnarray*}
	\ensuremath{\matrix{\mathrm{2}}{\mathrm{3}}{\mathrm{4}}{\mathrm{5}} \times \matrix{\mathrm{1}}{\mathrm{0}}{\mathrm{0}}{\mathrm{1}}\mathrel{=}\matrix{\mathrm{2} \times \mathrm{1}}{\mathrm{3} \times \mathrm{0}}{\mathrm{4} \times \mathrm{0}}{\mathrm{5} \times \mathrm{1}}\mathrel{=}\matrix{\mathrm{2}}{\mathrm{0}}{\mathrm{0}}{\mathrm{5}}}
\end{eqnarray*}
all of type \ensuremath{\mathrm{2}\to \mathrm{2}}.
Finally, for vectors \ensuremath{\Varid{v},\Varid{u}\mathbin{:}\Conid{A}\to \mathrm{1}},
\begin{eqnarray}
	\ensuremath{{\Varid{v}}\kr{\Varid{u}}\mathrel{=}\Varid{v} \times \Varid{u}}
	\label{eq:180225b}
\end{eqnarray}

\subsection{Measures}
As already said, the meaning of \emph{projection matrix} \ensuremath{\Varid{t}_{\Varid{d}}\hskip-0.5pt}, for \ensuremath{\Varid{d}}
a dimension of table \ensuremath{\Varid{t}}, is given by
\begin{eqnarray}
	\ensuremath{\Varid{v}\;\Varid{t}_{\Varid{d}}\hskip-0.5pt\;\Varid{i}\mathrel{=}\mathrm{1}} & \ensuremath{~\Leftrightarrow~} & \ensuremath{\lkp{\Varid{t}}{\Varid{d}}{\Varid{i}}\mathrel{=}\Varid{v}}
\end{eqnarray}
However, it does not seem a good idea to represent measure attribute
\ensuremath{\Varid{l\char95 extendedprice}} by projection matrix \ensuremath{\mean{\proj{\Varid{l}}{\Varid{extendedprice}}}\mathbin{:}\card{\Varid{l}}\to \R},
an infinite dimensional Boolean matrix.

Instead, LA offers the alternative
of ``internalizing" the real values of the dimension into a row vector, of type \ensuremath{\card{\Varid{l}}\to \mathrm{1}}
in the case of \ensuremath{\Varid{l\char95 extendedprice}}:\footnote{Again we abbreviate \ensuremath{\mathsf{\Varid{lineitem}}}
by \ensuremath{\mathsf{\Varid{l}}}, taking the converse so that the vector fits into one line of text.}
\begin{quote}
\ensuremath{\longlarrow{\mathrm{1}}{\conv{(\mea{\Varid{l}}{\Varid{extendedprice}})}}{\card{\Varid{l}}}} =
$\begin{bmatrix}
            2124.32
 \\        16994.56
 \\        32735.70
 \\         1902.10
 \\        42846.80
 \\        35707.22
 \\        44064.48
\end{bmatrix}$
\end{quote}
Also note that, to stress the difference between \emph{matrices} representing
\emph{dimension} attributes and \emph{vectors} representing \emph{measure}
attributes, we adopt a superscripted notation for the latter.

In general, a measure column (attribute) \ensuremath{\Varid{m}} of a source tabe \ensuremath{\mathsf{\Varid{t}}}
(e.g.\ \ensuremath{\Varid{m}\mathrel{=}\Varid{l\char95 extendedprice}}) is represented by a row vector of type \ensuremath{\card{\Varid{t}}\to \mathrm{1}}
whose cells contain the corresponding numeric data.\footnote{\emph{Measure}
columns are easy to spot in SQL data models by searching for \ensuremath{\Conid{DECIMAL}} SQL types.}
Superscripted notation \ensuremath{\mea{\Varid{t}}{\Varid{m}}\mathbin{:}\card{\Varid{t}}\to \mathrm{1}} distinguishes \emph{dimension}
matrices (e.g.\ \ensuremath{\proj{\Varid{t}}{\Varid{d}}}) from \emph{measure} vectors (e.g.\ \ensuremath{\mea{\Varid{t}}{\Varid{m}}}).
When needing to address the specific values of cells in measure vectors,
instead of (say) \ensuremath{\mathrm{1}\;\mea{\Varid{t}}{\Varid{m}}\;\Varid{i}}, for \ensuremath{\Varid{i}\;{\in}\;\card{\Varid{t}}}, we shall write
\ensuremath{\lkp{\Varid{t}}{\Varid{m}}{\Varid{i}}}.

\begin{figure}
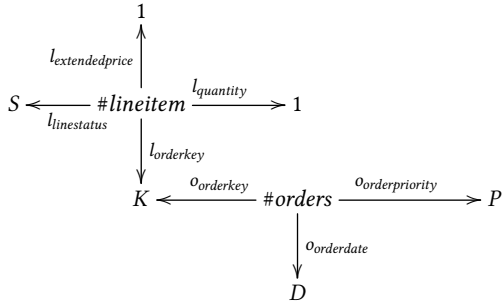

$
\myxym{
&
	\ensuremath{\mathrm{1}}
\\
	\ensuremath{\Conid{S}}
&
	\ensuremath{\card{\Varid{lineitem}}}
		\ar[l]^{\ensuremath{\proj{\Varid{l}}{\Varid{linestatus}}}}
		\ar[u]^{\ensuremath{\proj{\Varid{l}}{\Varid{extendedprice}}}}
		\ar[d]^{\ensuremath{\proj{\Varid{l}}{\Varid{orderkey}}}}
		\ar[r]^{\ensuremath{\proj{\Varid{l}}{\Varid{quantity}}}}
&
	\ensuremath{\mathrm{1}}
\\
&
	\ensuremath{\Conid{K}}
&
	\ensuremath{\card{\Varid{orders}}}
	\ar[rr]^{\ensuremath{\proj{\Varid{o}}{\Varid{orderpriority}}}}
	\ar[d]^{\ensuremath{\proj{\Varid{o}}{\Varid{orderdate}}}}
	\ar[l]_{\ensuremath{\proj{\Varid{o}}{\Varid{orderkey}}}}
&&
	\ensuremath{\Conid{P}}
\\
&
&
	\ensuremath{\Conid{D}}
&
}
$
\caption{Typed LA diagram representing the (simplified) schema of the TPC-H
database used as running example in the paper.\label{fig:180206a}}
\end{figure}

The typed LA diagram of Figure \ref{fig:180206a} depicts our definite model
of the schema of the TPC-H database used as running example in this paper.
Summarizing:
\begin{itemize}
\item 	\emph{measures} are (typed) \ensuremath{\Varid{vectors}}
\item	\emph{dimensions} are (typed) \ensuremath{\Varid{matrices}}.
\end{itemize}

Diagrams of this kind offer two main facilities. One is the ability to represent
data in a fully typed way. For instance, the information contained in table \ensuremath{\Varid{lineitem}} with respect
to quantities is captured by the column vector
\begin{eqnarray}
	\ensuremath{\Varid{v}\mathrel{=}({\proj{\Varid{l}}{\Varid{orderkey}}}\kr{\proj{\Varid{l}}{\Varid{linestatus}}}) \comp \conv{(\mea{\Varid{l}}{\Varid{quantity}})}}
	\label{eq:180214a}
\end{eqnarray}
of type	\ensuremath{\larrow{\mathrm{1}}{}{\Conid{K} \times \Conid{S}}}. To see what this means, we shall use
the following rules interfacing index-free and index-wise matrix notation,
where $M$ is an arbitrary matrix and $f$ and $g$ functional matrices:\footnote{These rules, expressed in the style of the Eindhoven quantifier calculus, are convenient shorthands for the corresponding instances of matrix composition (\ref{eq:120427a}). See Appendix \ref{sec:180220a} for an explanation of this style and notation.}
\begin{eqnarray}
	\mcell y{(\conv g \comp M\comp f)}x &\wider=& \mcell{(\ensuremath{\Varid{g}\;\Varid{y}})} M {(\ensuremath{\Varid{f}\;\Varid{x}})}
	\label{eq:120428d}
\\
	\mcell y{(f\comp M)}x &\wider=& \rcb\sum z {y=\ensuremath{\Varid{f}\;\Varid{z}}} {\mcell zMx}
	\label{eq:120428c}
\\
	\mcell y{(M\comp\conv f)}x &\wider=& \rcb\sum z {x=\ensuremath{\Varid{f}\;\Varid{z}}} {\mcell yMz}
	\label{eq:140321b}
\end{eqnarray}
Then (abbreviating \ensuremath{\proj{\Varid{l}}{\Varid{orderkey}}} to \ensuremath{\Varid{lky}} and \ensuremath{\proj{\Varid{l}}{\Varid{linestatus}}} to \ensuremath{\Varid{lst}}, for
saving space):
\begin{eqnarray*}
\start
	\ensuremath{\lkv{\Varid{v}}{\Varid{k},\Varid{s}}}
\just={ expanding shorthand notation \ensuremath{\lkv{\Varid{v}}{\Varid{k},\Varid{s}}} }
	\ensuremath{\mcell{(\Varid{k},\Varid{s})}{\Varid{v}}{\mathrm{1}}}
\just={ (\ref{eq:180214a}) and (\ref{eq:120428c}) }
	\ensuremath{\rcb{{\sum}}{\Varid{i}}{(\Varid{k},\Varid{s})\mathrel{=}({\Varid{lky}}\kr{\Varid{lst}})\;\Varid{i}}{\mcell{\Varid{i}}{\conv{(\mea{\Varid{l}}{\Varid{quantity}})}}{\mathrm{1}}}}
\just={ (\ref{eq:180214b}) and vector pointwise notation }
	\ensuremath{\rcb{{\sum}}{\Varid{i}}{\Varid{k}\mathrel{=}\Varid{lky}\;\Varid{i}\mathrel{\wedge}\Varid{s}\mathrel{=}\Varid{lst}\;\Varid{i}}{\lkv{\Varid{l\char95 quantity}}{\Varid{i}}}}
\end{eqnarray*}

\begin{figure}
\small
\begin{tabular}{cccc}
	&
\(
\ensuremath{\Varid{v}} =
\begin{array}{r|c}\ensuremath{\Conid{K} \times \Conid{S}}&{1}\\\hline
\hbox{(2723,F)}&0.0\\\hbox{(2723,O)}&4.0\\\hbox{(3392,F)}&1.0\\\hbox{(3392,O)}&3.0\\\hbox{(4354,F)}&5.0\\\hbox{(4354,O)}&9.0\\\hbox{(551,F)}&0.0\\\hbox{(551,O)}&1.0\\\hbox{(5699,F)}&7.0\\\hbox{(5699,O)}&0.0
\end{array}
\)
	&&
\(
Q=
\begin{array}{r|ccccc}&\rotatebox{90}{1992-07-30}&\rotatebox{90}{1994-09-30}&\rotatebox{90}{1995-05-30}&\rotatebox{90}{1995-10-06}&\rotatebox{90}{1995-10-28}\\\hline F&2&1&0&0&1\\O&0&1&1&1&1
\end{array}
\)
\\
	&
		(a)
	&&
		(b)
\end{tabular}
\caption{Vector \ensuremath{\Varid{v}\mathbin{:}\mathrm{1}\to (\Conid{K} \times \Conid{S})} (\ref{eq:180214a}) and query \ensuremath{\Conid{Q}\mathbin{:}\Conid{D}\to \Conid{S}} (\ref{eq:180215a}).\label{fig:180227a}}
\end{figure}

Vector \ensuremath{\Varid{v}} is depicted in Figure \ref{fig:180227a}a. The extra zeros correspond
to combinations of values in the attributes that could not be found in the
original data set. Strictly speaking, in a typed setting \emph{every pair}
of a cartesian product of types has to be taken into account.

The other feature is that data querying is performed simply by evaluating
paths of LA diagrams, just by composing the arrows, which are matrices. \emph{Converse}
is of capital importance to build these queries because it enables paths
that would otherwise not be available. For instance, the path
\begin{eqnarray}
	\ensuremath{\Conid{Q}\mathrel{=}\proj{\Varid{l}}{\Varid{linestatus}} \comp \conv{\proj{\Varid{l}}{\Varid{orderkey}}} \comp \proj{\Varid{o}}{\Varid{orderkey}} \comp \conv{\proj{\Varid{o}}{\Varid{orderdate}}}}
	\label{eq:180215a}
\end{eqnarray}
is the matrix of type \ensuremath{\Conid{S}\leftarrow \Conid{D}} depicted in Figure \ref{fig:180227a}b,
where \ensuremath{\cell{\Varid{s}}{\Conid{Q}}{\Varid{d}}} answers the query:
\begin{quote}\em
how many items are there with status \ensuremath{\Varid{s}} of orders issued on date \ensuremath{\Varid{d}}?
\end{quote}
This query is illustrative of two patterns that invariably turn up in querying diagrams,
as presented next.

\subsection{Joins and tabulations}
Let us consider two data sources \ensuremath{\Varid{p}} and \ensuremath{\Varid{t}} with attributes \ensuremath{\Conid{A}} and \ensuremath{\Conid{B}} as shown in the following diagram:
\begin{eqnarray}
\xymatrix{
&
	\ensuremath{\mathbin{\#}\Varid{p}}
	\ar@(ur,u)[]_{\ensuremath{\Conid{N}}}
	\ar[r]^{\ensuremath{\Varid{p}_{\Conid{A}}\hskip-0.5pt}}
	\ar[d]_{\ensuremath{\Varid{p}_{\Conid{B}}\hskip-0.5pt}}
	\ar[dl]_{\ensuremath{\Conid{X}}}
&
	\ensuremath{\Conid{A}}
	\ar[dl]^{\ensuremath{\Conid{Y}}}
\\
	\ensuremath{\mathbin{\#}\Varid{t}}
	\ar[r]_{\ensuremath{\Varid{t}_{\Conid{B}}\hskip-0.5pt}}
&
	\ensuremath{\Conid{B}}
	\ar@(d,r)[]_{\ensuremath{\Conid{M}}}
&
}
\end{eqnarray}
Let \ensuremath{\Conid{N}\mathbin{:}\card{\Varid{p}}\to \card{\Varid{p}}} and \ensuremath{\Conid{M}\mathbin{:}\Conid{B}\to \Conid{B}} be two arbitraty matrices of their type.
We refer to matrix
\ensuremath{\Conid{X}\mathbin{:}\card{\Varid{p}}\to \card{\Varid{t}}} defined by
\begin{eqnarray}
 	\ensuremath{\Conid{X}\mathrel{=}\conv{\Varid{t}_{\Conid{B}}\hskip-0.5pt} \comp \Conid{M} \comp \Varid{p}_{\Conid{B}}\hskip-0.5pt}
	\label{eq:180215b}
\end{eqnarray}
as a (generic) \emph{join}, and to matrix
\ensuremath{\Conid{Y}\mathbin{:}\Conid{A}\to \Conid{B}} defined by
\begin{eqnarray}
 	\ensuremath{\Conid{Y}\mathrel{=}\Varid{p}_{\Conid{B}}\hskip-0.5pt \comp \Conid{N} \comp \conv{\Varid{p}_{\Conid{A}}\hskip-0.5pt}}
	\label{eq:180217a}
\end{eqnarray}
as a (generic) \emph{tabulation}. Note (from their types) that joins are
matrices typed by \emph{indices} of two \emph{different} tables, while tabulations
are matrices typed by \emph{attributes} of the \emph{same} table.

Query \ensuremath{\Conid{Q}} (\ref{eq:180215a}) can be regarded as the composition of two tabulations in which \ensuremath{\Conid{N}\mathrel{=}\Varid{id}}. These are called \emph{counting} tabulations since the cells of the resulting matrices are all natural numbers. The corresponding situation in a \emph{join},
\ensuremath{\Conid{M}\mathrel{=}\Varid{id}} in (\ref{eq:180215b}),
is known as an \emph{equi-join}, e.g.
\small \begin{eqnarray*} \ensuremath{\conv{\proj{\Varid{l}}{\Varid{orderkey}}} \comp \proj{\Varid{o}}{\Varid{orderkey}}} =
\begin{array}{r|ccccc}&{1}&{2}&{3}&{4}&{5}\\\hline 1&0&0&0&1&0\\2&0&0&1&0&0\\3&1&0&0&0&0\\4&0&1&0&0&0\\5&0&0&0&0&1\\6&0&1&0&0&0\\7&1&0&0&0&0\\8&0&0&0&0&1
\end{array}
\end{eqnarray*} \normalsize
Equi-joins are always Boolean matrices that represent \emph{difunctional} relations
\cite{Ri48}.
Their pointwise meaning is the expected (make \ensuremath{\Conid{X}\mathrel{=}\conv{\proj{\Varid{l}}{\Varid{orderkey}}} \comp \proj{\Varid{o}}{\Varid{orderkey}}})
\begin{eqnarray*}
	\ensuremath{\cell{\Varid{j}}{\Conid{X}}{\Varid{i}}\mathrel{=}\lkp{\Varid{l}}{\Varid{orderkey}}{\Varid{j}}\mathrel{=}\lkp{\Varid{o}}{\Varid{orderkey}}{\Varid{i}}}
\end{eqnarray*}
recall rules (\ref{eq:120428d}) and (\ref{eq:180228b}).
That is, equi-join matrices correspond to the \emph{columnar-joins} of
\cite{Abadi2012-qq}, compare the example
\begin{center}
\begin{tikzpicture}[scale=.4]
\draw (0,0) rectangle (1,1) node[pos=.5] {\small 38};
\draw (0,1) rectangle (1,2) node[pos=.5] {\small 44};
\draw (0,2) rectangle (1,3) node[pos=.5] {\small 42};
\draw (0,3) rectangle (1,4) node[pos=.5] {\small 36};
\draw (0,4) rectangle (1,5) node[pos=.5] {\small 42};

\draw [white] (1.5,2) rectangle (2.5,3) node[pos=.5,black] { $\bowtie$};

\draw (3,0.5) rectangle (4,1.5) node[pos=.5] {\small 36};
\draw (3,1.5) rectangle (4,2.5) node[pos=.5] {\small 46};
\draw (3,2.5) rectangle (4,3.5) node[pos=.5] {\small 42};
\draw (3,3.5) rectangle (4,4.5) node[pos=.5] {\small 38};

\draw [white] (4.5,2) rectangle (5.5,3) node[pos=.5,black] { $=$};

\draw (6,0.5) rectangle (7,1.5) node[pos=.5] {\small 5};
\draw (6,1.5) rectangle (7,2.5) node[pos=.5] {\small 3};
\draw (6,2.5) rectangle (7,3.5) node[pos=.5] {\small 2};
\draw (6,3.5) rectangle (7,4.5) node[pos=.5] {\small 1};

\draw (8,0.5) rectangle (9,1.5) node[pos=.5] {\small 1};
\draw (8,1.5) rectangle (9,2.5) node[pos=.5] {\small 2};
\draw (8,2.5) rectangle (9,3.5) node[pos=.5] {\small 4};
\draw (8,3.5) rectangle (9,4.5) node[pos=.5] {\small 2};
\end{tikzpicture}
\end{center}
excerpted from \cite{Abadi2012-qq} with the same device encoded as a difunctional matrix:
\begin{quote}\small
\begin{tabular}{r|lllll}&1&2&3&4&5
\\\hline
   1	& 0	& 0	& 0	& 0	& 1
\\ 2	& 1	& 0	& 1	& 0	& 0
\\ 3	& 0	& 0	& 0	& 0	& 0
\\ 4	& 0	& 1	& 0	& 0	& 0
\end{tabular}
\end{quote}
Alternatively, (\ref{eq:180215a}) can be regarded as tabulation
\begin{eqnarray*}
	\ensuremath{\Conid{Q}\mathrel{=}\proj{\Varid{l}}{\Varid{linestatus}} \comp \Conid{M} \comp \conv{\proj{\Varid{o}}{\Varid{orderdate}}}}
\end{eqnarray*}
where \ensuremath{\Conid{M}} is the equi-join
\begin{eqnarray*}
	\ensuremath{\Conid{M}\mathrel{=}\conv{\proj{\Varid{l}}{\Varid{orderkey}}} \comp \proj{\Varid{o}}{\Varid{orderkey}}}.
\end{eqnarray*}

Other forms of joins and tabulations are obtained from
	(\ref{eq:180215b},\ref{eq:180217a})
by suitably instantiating matrices \ensuremath{\Conid{M}} and \ensuremath{\Conid{N}}.
For instance, let \ensuremath{\Conid{M}\mathrel{=}\mathopen\langle \leq \mathclose\rangle } in (\ref{eq:180215b}), the matrix encoding some ordering
on type \ensuremath{\Conid{B}}. Then \ensuremath{\Conid{X}\mathrel{=}\conv{\Varid{t}_{\Conid{B}}\hskip-0.5pt} \comp \mathopen\langle \leq \mathclose\rangle  \comp \Varid{p}_{\Conid{B}}\hskip-0.5pt} is the relational join
\begin{eqnarray*}
	\ensuremath{\cell{\Varid{j}}{\Conid{X}}{\Varid{i}}\mathrel{=}\lkp{\Varid{t}}{\Conid{B}}{\Varid{j}}\leq \lkp{\Varid{p}}{\Conid{B}}{\Varid{i}}}
\end{eqnarray*}
and so on and so forth.

Some explanation about the notation \ensuremath{\mathopen\langle \anonymous \mathclose\rangle } encoding predicates into
Boolean matrices is needed.
Given a \emph{binary} predicate $\ensuremath{\Varid{p}\mathbin{:}\Conid{B} \times \Conid{A}\to \mathbb B }$, denote by $\ensuremath{\mathopen\langle \Varid{p}\mathclose\rangle \mathbin{:}\Conid{B}\leftarrow \Conid{A}}$ the Boolean matrix which encodes $p$, that is,
\begin{eqnarray}
	\ensuremath{\Varid{b}\;\mathopen\langle \Varid{p}\mathclose\rangle \;\Varid{a}\mathrel{=}\mathbf{if}\;\Varid{p}\;(\Varid{b},\Varid{a})\;\mathbf{then}\;\mathrm{1}\;\mathbf{else}\;\mathrm{0}}
	\label{eq:160130a}
\end{eqnarray}
In case of a \emph{unary} predicate \ensuremath{\Varid{q}\mathbin{:}\Conid{A}\to \mathbb B }, \ensuremath{\mathopen\langle \Varid{q}\mathclose\rangle \mathbin{:}\mathrm{1}\leftarrow \Conid{A}} is
the Boolean vector such that:
\begin{eqnarray}
	\ensuremath{\mathrm{1}\;\mathopen\langle \Varid{q}\mathclose\rangle \;\Varid{a}\mathrel{=}\mathbf{if}\;\Varid{q}\;\Varid{a}\;\mathbf{then}\;\mathrm{1}\;\mathbf{else}\;\mathrm{0}}
	\label{eq:160102a}
\end{eqnarray}
We very often abbreviate the scalar \ensuremath{\mathrm{1}\;\mathopen\langle \Varid{q}\mathclose\rangle \;\Varid{a}} by \ensuremath{\lkv{\Varid{q}}{\Varid{a}}}.
By (\ref{eq:180225b}), the conjunction of two predicates is represented to the Hadamard
or Khatri-Rao product of the corresponding vectors:
	\ensuremath{\mathopen\langle \Varid{p}\mathrel{\wedge}\Varid{q}\mathclose\rangle \mathrel{=}{\mathopen\langle \Varid{p}\mathclose\rangle }\kr{\mathopen\langle \Varid{q}\mathclose\rangle }\mathrel{=}\mathopen\langle \Varid{p}\mathclose\rangle  \times \mathopen\langle \Varid{q}\mathclose\rangle }.

\subsection{Group-by's}
Group-by's are captured by tabulations (\ref{eq:180217a}) where matrix \ensuremath{\Conid{N}}
``absorbs" the measure data involved.

Let us see an example first, taken from the SQL query already given.
The aggregation is performed over the product
\ensuremath{\Varid{l\char95 quantity}\mathbin{*}\Varid{l\char95 extendedprice}}. In LA this corresponds to the pointwise product
of the two measure vectors, a particular case of the matrix Hadamard product (\ref{eq:180222a}).
So we can define 
\begin{eqnarray}
	\ensuremath{\rarrow{\card{\Varid{l}}}{\Varid{v}}{\mathrm{1}}\mathrel{=}\Varid{l}^{\Varid{extendedprice}} \times \Varid{l}^{\Varid{quantity}}}
\end{eqnarray}
and incorportate this vector in the part of the diagram involved in this query:
\begin{eqnarray*}
\xymatrix@R=3ex{
&
	\ensuremath{\card{\Varid{l}}}
		\ar[d]_{\ensuremath{\proj{\Varid{l}}{\Varid{orderkey}}}}
		\ar[r]^{\ensuremath{\Varid{v}}}
&
	\ensuremath{\mathrm{1}}
\\
&
	\ensuremath{\Conid{K}}
&
	\ensuremath{\card{\Varid{o}}}
	\ar[rr]^{\ensuremath{\proj{\Varid{o}}{\Varid{orderpriority}}}}
	\ar[d]_{\ensuremath{\proj{\Varid{o}}{\Varid{orderdate}}}}
	\ar[l]_{\ensuremath{\proj{\Varid{o}}{\Varid{orderkey}}}}
&&
	\ensuremath{\Conid{P}}
\\
&
&
	\ensuremath{\Conid{D}}
&
}
\end{eqnarray*}
By defining vector \ensuremath{\rarrow{\card{\Varid{o}}}{\Varid{u}}{\mathrm{1}}\mathrel{=}\Varid{v} \comp \conv{\proj{\Varid{l}}{\Varid{orderkey}}} \comp \proj{\Varid{o}}{\Varid{orderkey}}}
we get the \aspas{pairing wheel} (\ref{eq:180228e}):
\begin{eqnarray*}
\xymatrix{
&
	\ensuremath{\mathrm{1}}
\\
&
	\ensuremath{\card{\Varid{o}}}
		\ar[u]_{\ensuremath{\Varid{u}}}
		\ar[dr]^{\ensuremath{\proj{\Varid{o}}{\Varid{orderdate}}}}
		\ar[dl]_{\ensuremath{\proj{\Varid{o}}{\Varid{orderpriority}}}}
\\
	\ensuremath{\Conid{P}}
&&
	\ensuremath{\Conid{D}}
}
\end{eqnarray*}
As seen in appendix \ref{sec:18022b}, this can be expressed in several (isomorphic) ways.
One is the immediate
\begin{eqnarray}
\xymatrix@C=10em{
	\ensuremath{\Conid{D} \times \Conid{P}}
&
	1
		\ar[l]_-{\ensuremath{({\proj{\Varid{o}}{\Varid{orderdate}}}\kr{\proj{\Varid{o}}{\Varid{orderpriority}}}) \comp \conv{\Varid{u}}}}
}
	\label{eq:180222f}
\end{eqnarray}
--- cf.\ (\ref{eq:180222d}) --- yielding the tensor presentation\footnote{Recall that matrices or vectors  of higher-rank such as this are normally known as \emph{tensors} \cite{STPYF08}.}
\small
\begin{eqnarray*}
\begin{array}{r|r}\ensuremath{\Conid{D} \times \Conid{P}}&{1}\\\hline
  \hbox{(1992-07-30,2-HIGH)}&285793.8
\\\hbox{(1994-09-30,2-HIGH)}&16994.56
\\\hbox{(1994-09-30,3-MEDIUM)}&8497.28
\\\hbox{(1995-10-06,2-HIGH)}&195655.0
\\\hbox{(1995-10-28,3-MEDIUM)}&64353.3
\end{array}
\end{eqnarray*}
\normalsize
for the sample data of our running example, removing zero entries
for space economy. Now, by (\ref{eq:180222e}) in appendix \ref{sec:18022b} we can also
express the same information by
\begin{eqnarray*}
\xymatrix@C=6em{
	\ensuremath{\Conid{D}}
&
	\ensuremath{\card{\Varid{o}}}
		\ar[l]_-{\ensuremath{{\proj{\Varid{o}}{\Varid{orderdate}}}\kr{\Varid{u}}}}
&
	\ensuremath{\Conid{P}}
		\ar[l]_-{\ensuremath{\conv{\proj{\Varid{o}}{\Varid{orderpriority}}}}}
}
\end{eqnarray*}
or by
\begin{eqnarray}
\xymatrix@C=6em{
	\ensuremath{\Conid{P}}
&
	\ensuremath{\card{\Varid{o}}}
		\ar[l]_-{\ensuremath{{\proj{\Varid{o}}{\Varid{orderpriority}}}\kr{\Varid{u}}}}
&
	\ensuremath{\Conid{D}}
		\ar[l]_-{\ensuremath{\conv{\proj{\Varid{o}}{\Varid{orderdate}}}}}
}
	\label{eq:180228c}
\end{eqnarray}
in this case yielding the same data in tabulated format:
\small
\begin{eqnarray*}
\begin{array}{r|ccccc}&\rotatebox{90}{1992-07-30}&\rotatebox{90}{1994-09-30}&\rotatebox{90}{1995-05-30}&\rotatebox{90}{1995-10-06}&\rotatebox{90}{1995-10-28}\\\hline \hbox{2-HIGH}&285793.8&0.0&16994.5&8497.2&0.0\\\hbox{3-MEDIUM}&0.0&195655.0&0.0&0.0&64353.3
\end{array}
\end{eqnarray*}

\normalsize
Path (\ref{eq:180228c})  can be further re-written, by (\ref{eq:160101a}), into the tabulation format
(\ref{eq:180217a})
\begin{eqnarray*}
\xymatrix@C=4em{
	\ensuremath{\Conid{P}}
&
	\ensuremath{\card{\Varid{o}}}
		\ar[l]_-{\ensuremath{\proj{\Varid{o}}{\Varid{orderpriority}}}}
&
	\ensuremath{\card{\Varid{o}}}
		\ar[l]^-{\ensuremath{\underbrace{{\Varid{id}}\kr{\Varid{u}}}_{\Conid{N}}}}
&
	\ensuremath{\Conid{D}}
		\ar[l]_-{\ensuremath{\conv{\proj{\Varid{o}}{\Varid{orderdate}}}}}
}
\end{eqnarray*}
for \ensuremath{\Conid{N}\mathrel{=}{\Varid{id}}\kr{\Varid{u}}}, the diagonal matrix representing vector \ensuremath{\Varid{u}}.
That is, non-zeros can only be found in the \emph{diagonal} of this square matrix,
cf:
\begin{eqnarray*}
\start
	\ensuremath{\Varid{j}\;({\Varid{id}}\kr{\Varid{u}})\;\Varid{i}} \wider= \ensuremath{(\Varid{j}\;\Varid{id}\;\Varid{i}) \times \lkv{\Varid{u}}{\Varid{i}}}
\just\equiv{ Khatri-Rao (\ref{eq:130129b}) ; \ensuremath{\Varid{id}\;\Varid{x}\mathrel{=}\Varid{x}} }
	\ensuremath{\Varid{j}\;({\Varid{id}}\kr{\Varid{u}})\;\Varid{i}} \wider= \ensuremath{(\Varid{j}\mathrel{=}\Varid{i}) \times \lkv{\Varid{u}}{\Varid{i}}}
\just\equiv{ pointwise LA rule (\ref{eq:160130a}) }
	\ensuremath{\Varid{j}\;({\Varid{id}}\kr{\Varid{u}})\;\Varid{i}} \wider= \ensuremath{\mathbf{if}\;\Varid{j}\mathrel{=}\Varid{i}\;\mathbf{then}\;\lkv{\Varid{u}}{\Varid{i}}\;\mathbf{else}\;\mathrm{0}}
\end{eqnarray*}

Altogther, query
\begin{eqnarray*}
	\ensuremath{\larrow{\Conid{D}}{\Conid{Q}}{\Conid{P}}\mathrel{=}\proj{\Varid{o}}{\Varid{orderpriority}} \comp \underbrace{{\Varid{id}}\kr{\Varid{u}}}_{\Conid{N}} \comp \conv{\proj{\Varid{o}}{\Varid{orderdate}}}}
\end{eqnarray*}
evaluates in the following way:
for each \ensuremath{\Varid{p}\;{\in}\;\Conid{P}} and \ensuremath{\Varid{d}\;{\in}\;\Conid{D}}, it finds those records \ensuremath{\Varid{i}\;{\in}\;\card{\Varid{o}}} with \ensuremath{\Varid{p}} and \ensuremath{\Varid{d}} as
attributes and adds up prices \ensuremath{\lkv{\Varid{u}}{\Varid{i}}}, cf.
\begin{eqnarray*}
\start
	\ensuremath{\Varid{p}\;\Conid{Q}\;\Varid{d}}
\just={ rule (\ref{eq:120428c}) etc }
	\ensuremath{\rcb{{\sum}}{\Varid{j},\Varid{i}}{\Varid{p}\mathrel{=}\proj{\Varid{o}}{\Varid{orderpriority}}\;\Varid{j}\mathrel{\wedge}\Varid{d}\mathrel{=}\proj{\Varid{o}}{\Varid{orderdate}}\;\Varid{i}}{\Varid{j}\;\Conid{N}\;\Varid{i}}}
\just={ since \ensuremath{\cell{\Varid{j}}{\Conid{N}}{\Varid{i}}} \wider= \ensuremath{\mathbf{if}\;\Varid{j}\mathrel{=}\Varid{i}\;\mathbf{then}\;\lkv{\Varid{u}}{\Varid{i}}\;\mathbf{else}\;\mathrm{0}}}
	\ensuremath{\rcb{{\sum}}{\Varid{i}}{\Varid{p}\mathrel{=}\proj{\Varid{o}}{\Varid{orderpriority}}\;\Varid{i}\mathrel{\wedge}\Varid{d}\mathrel{=}\proj{\Varid{o}}{\Varid{orderdate}}\;\Varid{i}}{\lkv{\Varid{u}}{\Varid{i}}}}
\end{eqnarray*}
In summary: given
\begin{eqnarray*}
\xymatrix@R=3em{
	\ensuremath{\mathrm{1}}
&
	\ensuremath{\mathbin{\#}\Varid{p}}
	\ar@(ur,u)[]_{\ensuremath{\Conid{N}\mathrel{=}{\Varid{id}}\kr{\Varid{u}}}}
	\ar[r]^{\ensuremath{\Varid{p}_{\Conid{A}}\hskip-0.5pt}}
	\ar[d]_{\ensuremath{\Varid{p}_{\Conid{B}}\hskip-0.5pt}}
	\ar[l]_{\ensuremath{\Varid{u}}}
&
	\ensuremath{\Conid{A}}
	\ar[dl]^{\ensuremath{\Conid{Y}}}
\\
&
	\ensuremath{\Conid{B}}
&
}
\end{eqnarray*}
there are three equivalent ways of expressing \emph{tabulation} \ensuremath{\Conid{Y}}:
\begin{eqnarray}
\begin{array}{rcl}
 	\ensuremath{\Conid{Y}} &=& \ensuremath{\Varid{p}_{\Conid{B}}\hskip-0.5pt \comp ({\Varid{id}}\kr{\Varid{u}}) \comp \conv{\Varid{p}_{\Conid{A}}\hskip-0.5pt}}
\\ &=&
 	\ensuremath{({\Varid{p}_{\Conid{B}}\hskip-0.5pt}\kr{\Varid{u}}) \comp \conv{\Varid{p}_{\Conid{A}}\hskip-0.5pt}}
\\ &=&
 	\ensuremath{\Varid{p}_{\Conid{B}}\hskip-0.5pt \comp \conv{({\Varid{p}_{\Conid{A}}\hskip-0.5pt}\kr{\Varid{u}})}}.
\end{array}
	\label{eq:180225a}
\end{eqnarray}
That is, a \emph{measure} or \emph{condition} captured by a vector \ensuremath{\Varid{u}} can
be incorporated in a tabulation by multiplying (in the Khatri-Rao sense)
any of its ``legs" \ensuremath{\Varid{p}_{\Conid{B}}\hskip-0.5pt} or \ensuremath{\Varid{p}_{\Conid{A}}\hskip-0.5pt} by \ensuremath{\Varid{u}}. The second and
third alternatives are handy in the sense that they allow for simplifying
LA paths, as we shall soon see.

In the following sections we shall see how to evaluate typed LA paths
on top of an LA kernel implementing the operations involved (Khatri-Rao, etc),
making it possible to benchmark our approach.
Relying on law (\ref{eq:180222b}) of the appendix,
\begin{eqnarray*}
	\ensuremath{\Conid{P} \comp \conv{\Varid{v}}\mathrel{=}({\Varid{v}}\kr{\Conid{P}}) \comp \conv{{\bang}}}
\end{eqnarray*}
tabulations of type \ensuremath{\Conid{B}\leftarrow \Conid{A}} are rendered in the type \ensuremath{\Conid{B} \times \Conid{A}\leftarrow \mathrm{1}}, that is, in the
same columnar format as (\ref{eq:180222f}). In the LA kernel,
\ensuremath{( \comp \conv{{\bang}})} is implemented by a convenient operation that adds matrices row-wise.
For instance, in the case of (\ref{eq:180222f}), the implemented LA script 
is equivalent to the expression:
\begin{eqnarray*}
	\ensuremath{({\Varid{u}}\kr{({\proj{\Varid{o}}{\Varid{orderdate}}}\kr{\proj{\Varid{o}}{\Varid{orderpriority}}})}) \comp \conv{{\bang}}}
\end{eqnarray*}

We detail the query conversion into a LA script in section \ref{sec:180222g}, while implementation details are in section 6.
Before that, we address an important feature of the LA approach
proposed in this paper: incrementality (i.e.\ differential querying).

%% file: sections/4-incremental.tex
\section{Incremental querying}\label{sec:180301a}
One advantage of the typed LA approach to data querying is \emph{incrementality}.
Suppose that, having evaluated a query \ensuremath{\Conid{Q}\;(\Conid{M})} involving some given data encoded into
matrix \ensuremath{\Conid{M}}, data envolves so that \ensuremath{\Conid{M}} becomes \ensuremath{\Conid{M'}}.

Because \ensuremath{\Conid{M}} and \ensuremath{\Conid{M'}} are matrices, we can calculate their difference, \ensuremath{\delta\;\Conid{M}\mathrel{=}\Conid{M'}\mathbin{-}\Conid{M}}. Then, instead of calculating \ensuremath{\Conid{Q}\;(\Conid{M'})} we can add to the already calculated
\ensuremath{\Conid{Q}\;(\Conid{M})} the corresponding \emph{differential} query \ensuremath{\Conid{Q'}\;(\delta\;\Conid{M})}, cf:
\begin{eqnarray}
	\ensuremath{\Conid{Q}\;(\Conid{M}\mathbin{+}\delta\;\Conid{M})\mathrel{=}\Conid{Q}\;(\Conid{M})\mathbin{+}\Conid{Q'}\;(\delta\;\Conid{M})}
	\label{eq:180228d}
\end{eqnarray}
The derivation of \ensuremath{\Conid{Q'}} can always be performed
thanks to the \emph{linearity} of the LA operators, notably of composition,
\begin{eqnarray*}
	M\comp (N + P) &=& M\comp N + M\comp P
\\
	(N + P)\comp M &=& N\comp M + P\comp M
\end{eqnarray*}
of converse,
\begin{eqnarray*}
	\ensuremath{\conv{(\Conid{M}\mathbin{+}\Conid{N})}\mathrel{=}\conv{\Conid{M}}\mathbin{+}\conv{\Conid{N}}}
\end{eqnarray*}
and of the Khatri-Rao product:
\begin{eqnarray*}
	\ensuremath{{(\Conid{M}\mathbin{+}\Conid{N})}\kr{\Conid{P}}\mathrel{=}{\Conid{M}}\kr{\Conid{P}}\mathbin{+}{\Conid{N}}\kr{\Conid{P}}}
\\
	\ensuremath{{\Conid{P}}\kr{(\Conid{M}\mathbin{+}\Conid{N})}\mathrel{=}{\Conid{P}}\kr{\Conid{M}}\mathbin{+}{\Conid{P}}\kr{\Conid{N}}}
\end{eqnarray*}
In the case of queries involving several parameters,
for instance three parameters such as in
\begin{quote}
	\ensuremath{({\proj{\Varid{o}}{\Varid{orderdate}}}\kr{\proj{\Varid{o}}{\Varid{orderpriority}}}) \comp \conv{\Varid{u}}}
\end{quote}
one performs accordingly for all the parameters.

The main advantage is that, assuming \ensuremath{\Conid{Q}\;(\Conid{M})} already
calculated in (\ref{eq:180228d}), \ensuremath{\Conid{Q'}\;(\delta\;\Conid{M})} costs far less than 
	\ensuremath{\Conid{Q}\;(\Conid{M}\mathbin{+}\delta\;\Conid{M})}
due to the reduced size of \ensuremath{\delta\;\Conid{M}} compared to \ensuremath{\Conid{M}}.

%% file: sections/5-SQL-LA.tex
\section{From SQL to typed LA scripts}
\label{sec:180222g}
We are ready to explain the process of converting a realistic SQL query into
its equivalent typed LA script. Query 3 in the TPC-H benchmark suite \cite{TPC17}
is taken as example.
\begin{hscode}\SaveRestoreHook
\column{B}{@{}>{\hspre}l<{\hspost}@{}}%
\column{4}{@{}>{\hspre}l<{\hspost}@{}}%
\column{9}{@{}>{\hspre}l<{\hspost}@{}}%
\column{E}{@{}>{\hspre}l<{\hspost}@{}}%
\>[4]{} \sql{select} \;{}\<[E]%
\\
\>[4]{}\hsindent{5}{}\<[9]%
\>[9]{}\Varid{l\char95 orderkey},\Varid{o\char95 orderdate},\Varid{o\char95 shippriority};{}\<[E]%
\\
\>[4]{}\hsindent{5}{}\<[9]%
\>[9]{}\Varid{sum}\;(\Varid{l\char95 extendedprice}\mathbin{*}(\mathrm{1}\mathbin{-}\Varid{l\char95 discount}))\;\Varid{as}\;\Varid{revenue}{}\<[E]%
\\
\>[4]{} \sql{from} \;{}\<[E]%
\\
\>[4]{}\hsindent{5}{}\<[9]%
\>[9]{}\Varid{orders},\Varid{customer},\Varid{lineitem}{}\<[E]%
\\
\>[4]{}\mathbf{where}{}\<[E]%
\\
\>[4]{}\hsindent{5}{}\<[9]%
\>[9]{}\Varid{c\char95 mktsegment}\mathrel{=}\text{\ttfamily 'MACHINERY'}{}\<[E]%
\\
\>[4]{}\hsindent{5}{}\<[9]%
\>[9]{}\Varid{and}\;\Varid{c\char95 custkey}\mathrel{=}\Varid{o\char95 custkey}{}\<[E]%
\\
\>[4]{}\hsindent{5}{}\<[9]%
\>[9]{}\Varid{and}\;\Varid{l\char95 orderkey}\mathrel{=}\Varid{o\char95 orderkey}{}\<[E]%
\\
\>[4]{}\hsindent{5}{}\<[9]%
\>[9]{}\Varid{and}\;\Varid{o\char95 orderdate}\mathbin{<}\Varid{date}\;\text{\ttfamily '1995-03-10'}{}\<[E]%
\\
\>[4]{}\hsindent{5}{}\<[9]%
\>[9]{}\Varid{and}\;\Varid{l\char95 shipdate}\mathbin{>}\Varid{date}\;\text{\ttfamily '1995-03-10'}{}\<[E]%
\\
\>[4]{} \sql{group\ by} \;{}\<[E]%
\\
\>[4]{}\hsindent{5}{}\<[9]%
\>[9]{}\Varid{l\char95 orderkey},\Varid{o\char95 orderdate},\Varid{o\char95 shippriority}{}\<[E]%
\\
\>[4]{} \sql{order\ by} \;{}\<[E]%
\\
\>[4]{}\hsindent{5}{}\<[9]%
\>[9]{}\Varid{revenue}\;\Varid{desc},\Varid{o\char95 orderdate};{}\<[E]%
\ColumnHook
\end{hscode}\resethooks
The type diagram given before needs to be extended with
a new table --- \ensuremath{\Varid{customer}}, abbreviated to \ensuremath{\Varid{c}} --- with the attributes (projection
functions) as shown in the diagram:
\begin{eqnarray*}
\xymatrix@C=5em@R=2em{
&
&
	\ensuremath{\Conid{P}}
&
&
\\
&
	\ensuremath{\Conid{K}}
&
	\ensuremath{\mathbin{\#}\Varid{o}}
	\ar[u]_{\ensuremath{\Varid{o}_{\Varid{shippriority}}\hskip-0.5pt}}
	\ar[r]_{\ensuremath{\Varid{o}_{\Varid{custkey}}\hskip-0.5pt}}
	\ar[l]^{\ensuremath{\Varid{o}_{\Varid{orderkey}}\hskip-0.5pt}}
	\ar[d]^{\ensuremath{\Varid{o}_{\Varid{orderdate}}\hskip-0.5pt}}
&
	\ensuremath{\Conid{C}}
&
\\
&
	\ensuremath{\mathbin{\#}\Varid{l}}
		\ar[u]^{\ensuremath{\Varid{l}_{\Varid{orderkey}}\hskip-0.5pt}}
		\ar[r]_{\ensuremath{\Varid{l}_{\Varid{shipdate}}\hskip-0.5pt}}
		\ar[d]_{\ensuremath{\Varid{l}_{\Varid{extendedprice}}\hskip-0.5pt}}
		\ar[d]^{\ensuremath{\Varid{l}_{\Varid{discount}}\hskip-0.5pt}}
&
	\ensuremath{\Conid{D}}
&
	\ensuremath{\mathbin{\#}\Varid{c}}
		\ar[u]_{\ensuremath{\Varid{c}_{\Varid{custkey}}\hskip-0.5pt}}
		\ar[d]_{\ensuremath{\Varid{c}_{\Varid{mktsegment}}\hskip-0.5pt}}
&
\\
&
	\ensuremath{\mathrm{1}}
&
&
	\ensuremath{\Conid{S}}
}
\end{eqnarray*}

Looking at the group-by clause,
\begin{quote}
   \ensuremath{ \sql{group\ by} \;\Varid{l\char95 orderkey},\Varid{o\char95 orderdate},\Varid{o\char95 shippriority}}
\end{quote}
we aim at building a path of type \ensuremath{\Conid{K} \times (\Conid{D} \times \Conid{P})\leftarrow \mathrm{1}} (column vector) or of any equivalent
(isomorphic) type.
We start by encoding the predicates specified in the where-clause of the query
by Boolean vectors
\begin{quote}
\ensuremath{\Varid{u}\mathbin{:}\mathbin{\#}\Varid{l}\to \mathrm{1}}
\\
\ensuremath{\Varid{v}\mathbin{:}\mathbin{\#}\Varid{o}\to \mathrm{1}}
\end{quote}
defined by:
\begin{quote}
\ensuremath{\Varid{u}\;[\mskip1.5mu \Varid{i}\mskip1.5mu]\mathrel{=}\lkp{\Varid{l}}{\Varid{shipdate}}{\Varid{i}}\mathbin{>}\text{\ttfamily '1995-03-10'}}
\\
\ensuremath{\Varid{v}\;[\mskip1.5mu \Varid{j}\mskip1.5mu]\mathrel{=}\lkp{\Varid{o}}{\Varid{orderdate}}{\Varid{j}}\mathbin{<}\text{\ttfamily '1995-03-10'}}
\end{quote}
recall (\ref{eq:160102a}). Data value {\small\tt MACHINERY} is captured by the constant
function \ensuremath{\rarrow{\mathrm{1}}{\kons{\texttt{MACHINERY}}}{\Conid{S}}} such that \ensuremath{\kons{\texttt{MACHINERY}}\;\mathrm{1}} = {\small\tt MACHINERY}.\footnote{Constant functions of this kind, also called \emph{points}, are a standard algebraic way of describing specific data values.}
Moreover, clauses
\begin{hscode}\SaveRestoreHook
\column{B}{@{}>{\hspre}l<{\hspost}@{}}%
\column{E}{@{}>{\hspre}l<{\hspost}@{}}%
\>[B]{}\Varid{c\char95 mktsegment}\mathrel{=}\text{\ttfamily 'MACHINERY'}{}\<[E]%
\\
\>[B]{}\Varid{and}\;\Varid{c\char95 custkey}\mathrel{=}\Varid{o\char95 custkey}{}\<[E]%
\ColumnHook
\end{hscode}\resethooks
amount to Boolean path (vector)
\begin{eqnarray*}
z =
\xymatrix@C=09ex{
	1
&
	\ensuremath{\Conid{S}}
		\ar[l]_{\ensuremath{\conv{\kons{\texttt{MACHINERY}}}}}
&
	\ensuremath{\mathbin{\#}\Varid{c}}
		\ar[l]_{\ensuremath{\Varid{c}_{\Varid{mktsegment}}\hskip-0.5pt}}
&
	\ensuremath{\Conid{C}}
		\ar[l]_{\ensuremath{\conv{\Varid{c}_{\Varid{custkey}}\hskip-0.5pt}}}
&
	\ensuremath{\mathbin{\#}\Varid{o}}
		\ar[l]_{\ensuremath{\Varid{o}_{\Varid{custkey}}\hskip-0.5pt}}
}
\end{eqnarray*}
which \emph{counts} how many customers exhibit the specified market segment:
\begin{hscode}\SaveRestoreHook
\column{B}{@{}>{\hspre}l<{\hspost}@{}}%
\column{10}{@{}>{\hspre}l<{\hspost}@{}}%
\column{14}{@{}>{\hspre}l<{\hspost}@{}}%
\column{E}{@{}>{\hspre}l<{\hspost}@{}}%
\>[B]{}\Varid{z}\;[\mskip1.5mu \Varid{k}\mskip1.5mu]\mathrel{=}\rcb{{\sum}}{\Varid{i}}{\Varid{sel}\;\Varid{i}}{\mathrm{1}}{}\<[E]%
\\
\>[B]{}\hsindent{10}{}\<[10]%
\>[10]{}\mathbf{where}\;\Varid{sel}\;\Varid{i}\mathrel{=}{}\<[E]%
\\
\>[10]{}\hsindent{4}{}\<[14]%
\>[14]{}\lkp{\Varid{c}}{\Varid{custkey}}{\Varid{i}}\mathrel{=}\lkp{\Varid{o}}{\Varid{custkey}}{\Varid{k}}\mathrel{\wedge}{}\<[E]%
\\
\>[10]{}\hsindent{4}{}\<[14]%
\>[14]{}\lkp{\Varid{c}}{\Varid{mktsegment}}{\Varid{i}}\mathrel{=}\texttt{MACHINERY}{}\<[E]%
\ColumnHook
\end{hscode}\resethooks
Finally, we also define \emph{measure} vector \ensuremath{\rarrow{\card{\Varid{l}}}{\Varid{revenue}}{\mathrm{1}}} in the expected way:
\begin{eqnarray}
	\ensuremath{\Varid{revenue}\mathrel{=}\Varid{l}_{\Varid{extendedprice}}\hskip-0.5pt \times ({\bang}\mathbin{-}\Varid{l}_{\Varid{discount}}\hskip-0.5pt)}
	\label{eq:151215a}
\end{eqnarray}

Altogether, we obtain:
\begin{eqnarray*}
\start
\xymatrix@C=5em@R=2em{
&
&
	\ensuremath{\Conid{P}}
&
&
\\
&
	\ensuremath{\Conid{K}}
&
	\ensuremath{\mathbin{\#}\Varid{o}}
	\ar[u]_{\ensuremath{\Varid{o}_{\Varid{shippriority}}\hskip-0.5pt}}
	\ar[r]^{\ensuremath{\Varid{o}_{\Varid{orderdate}}\hskip-0.5pt}}
	\ar[l]^{\ensuremath{\Varid{o}_{\Varid{orderkey}}\hskip-0.5pt}}
	\ar[d]^{\ensuremath{\Varid{v} \times \Varid{z}}}
&
	\ensuremath{\Conid{D}}
&
\\
	\ensuremath{\mathrm{1}}
&
	\ensuremath{\mathbin{\#}\Varid{l}}
		\ar[u]^{\ensuremath{\Varid{l}_{\Varid{orderkey}}\hskip-0.5pt}}
		\ar[l]_{\ensuremath{\Varid{u}}}
		\ar[r]_{\ensuremath{\Varid{revenue}}}
&
	\ensuremath{\mathrm{1}}
&
}
\end{eqnarray*}
By (\ref{eq:180225a}) we define
\begin{quote}
\ensuremath{\Varid{o}_{\Varid{orderdate'}}\hskip-0.5pt\mathrel{=}{\Varid{o}_{\Varid{orderdate}}\hskip-0.5pt}\kr{(\Varid{v} \times \Varid{z})}}
\\
\ensuremath{\Varid{l}_{\Varid{orderkey'}}\hskip-0.5pt\mathrel{=}{\Varid{l}_{\Varid{orderkey}}\hskip-0.5pt}\kr{\Varid{u}}}
\end{quote}
we merge predicates with projections, which thus become \emph{partial} functions:
\begin{eqnarray*}
\start
\xymatrix@C=5em@R=2em{
&
	\ensuremath{\Conid{P}}
&
&
\\
	\ensuremath{\Conid{K}}
&
	\ensuremath{\mathbin{\#}\Varid{o}}
	\ar[u]^{\ensuremath{\Varid{o}_{\Varid{shippriority}}\hskip-0.5pt}}
	\ar[r]^{\ensuremath{\Varid{o}_{\Varid{orderdate'}}\hskip-0.5pt}}
	\ar[l]^{\ensuremath{\Varid{o}_{\Varid{orderkey}}\hskip-0.5pt}}
&
	\ensuremath{\Conid{D}}
&
\\
	\ensuremath{\mathbin{\#}\Varid{l}}
		\ar[u]^{\ensuremath{\Varid{l}_{\Varid{orderkey'}}\hskip-0.5pt}}
		\ar[r]_{\ensuremath{\Varid{revenue}}}
&
	\ensuremath{\mathrm{1}}
&
}
\end{eqnarray*}
Note how vector
\begin{eqnarray*}
	\ensuremath{\Varid{r}\mathrel{=}\Varid{revenue} \comp \conv{\Varid{l}_{\Varid{orderkey'}}\hskip-0.5pt}}
\end{eqnarray*}
computes measure \ensuremath{\Varid{revenue}} (filtered by \ensuremath{\Varid{u}}), for each key \ensuremath{\Varid{k}}:
\begin{eqnarray*}
	\ensuremath{\lkv{\Varid{r}}{\Varid{k}}\mathrel{=}\rcb{{\sum}}{\Varid{i}}{\Varid{k}\mathrel{=}\Varid{l}_{\Varid{orderkey}}\hskip-0.5pt\mathrel{\wedge}\lkv{\Varid{u}}{\Varid{i}}}{\lkv{\Varid{revenue}}{\Varid{i}}}}
\end{eqnarray*}
We can merge \ensuremath{\Varid{r}} with \ensuremath{\Varid{o}_{\Varid{orderkey}}\hskip-0.5pt},
\begin{quote}
\ensuremath{\Varid{o}_{\Varid{orderkey'}}\hskip-0.5pt\mathrel{=}\conv{({\Varid{r}}\kr{\conv{\Varid{o}_{\Varid{orderkey}}\hskip-0.5pt}})}}
\end{quote}
to obtain as outcome the expected `pairing wheel':
\begin{eqnarray*}
\xymatrix@C=2ex@R=3ex{
	\ensuremath{\Conid{P}}
&&
	\ensuremath{\Conid{D}}
\\
&
	\ensuremath{\card{\Varid{o}}}
		\ar[d]_{\ensuremath{\Varid{o}_{\Varid{orderkey'}}\hskip-0.5pt}}
		\ar[ur]_{\ensuremath{\proj{\Varid{o}}{\Varid{orderdate'}}}}
		\ar[ul]^{\ensuremath{\proj{\Varid{o}}{\Varid{orderpriority}}}}
\\
&
	\ensuremath{\Conid{K}}
}
\end{eqnarray*}
Finally, putting everything together:
\begin{hscode}\SaveRestoreHook
\column{B}{@{}>{\hspre}l<{\hspost}@{}}%
\column{3}{@{}>{\hspre}l<{\hspost}@{}}%
\column{5}{@{}>{\hspre}l<{\hspost}@{}}%
\column{E}{@{}>{\hspre}l<{\hspost}@{}}%
\>[B]{}\larrow{\Conid{K}}{\Conid{Q}}{\Conid{P} \times \Conid{D}}\mathrel{=}\Conid{H} \comp \Conid{X}{}\<[E]%
\\
\>[B]{}\hsindent{3}{}\<[3]%
\>[3]{}\mathbf{where}{}\<[E]%
\\
\>[3]{}\hsindent{2}{}\<[5]%
\>[5]{}\Conid{H}\mathrel{=}{\proj{\Varid{o}}{\Varid{orderpriority}}}\kr{({({\Varid{o}_{\Varid{orderdate}}\hskip-0.5pt}\kr{\Varid{v}})}\kr{\Varid{z}})}{}\<[E]%
\\
\>[3]{}\hsindent{2}{}\<[5]%
\>[5]{}\Conid{X}\mathrel{=}({(\Varid{revenue} \comp \Varid{l}_{\Varid{orderkey'}}\hskip-0.5pt)}\kr{\conv{\Varid{o}_{\Varid{orderkey}}\hskip-0.5pt}}){}\<[E]%
\ColumnHook
\end{hscode}\resethooks

The actual implementation of \ensuremath{\Conid{Q}} over the developed LA kernel offers the output
in isomorphic type \ensuremath{\Conid{K} \times (\Conid{P} \times \Conid{D})\leftarrow \mathrm{1}} using the rotating isomorphism \ensuremath{\alpha } of
appendix \ref{sec:18022b}. Written in a simple, single-assignment language defined
for guiding the implementation process, \ensuremath{\Conid{Q}} reads like this:
\begin{quote}\small
\begin{tabbing}\ttfamily
~v~\char61{}~filter\char40{}~o\char95{}orderdate~\char60{}~\char39{}1995\char45{}03\char45{}10\char39{}~\char41{}\\
\ttfamily ~B~\char61{}~krao\char40{}~v\char44{}~o\char95{}orderdate~\char41{}\\
\ttfamily ~C~\char61{}~filter\char40{}~c\char95{}mktsegment~\char61{}~\char39{}MACHINERY\char39{}~\char41{}\\
\ttfamily ~u~\char61{}~filter\char40{}~l\char95{}shipdate~\char62{}~\char39{}1995\char45{}03\char45{}10\char39{}~\char41{}\\
\ttfamily ~z~\char61{}~dot\char40{}~C\char44{}~o\char95{}custkey~\char41{}\\
\ttfamily ~F~\char61{}~krao\char40{}~l\char95{}orderkey\char44{}~u~\char41{}\\
\ttfamily ~G~\char61{}~krao\char40{}~B\char44{}~z~\char41{}\\
\ttfamily ~H~\char61{}~krao\char40{}~G\char44{}~o\char95{}shippriority~\char41{}\\
\ttfamily ~I~\char61{}~dot\char40{}~H\char44{}~l\char95{}orderkey~\char41{}\\
\ttfamily ~J~\char61{}~krao\char40{}~F\char44{}~I~\char41{}\\
\ttfamily ~K~\char61{}~lift\char40{}~l\char95{}extendedprice~\char42{}~\char40{}1\char45{}l\char95{}discount\char41{}~\char41{}\\
\ttfamily ~L~\char61{}~krao\char40{}~J\char44{}~K~\char41{}\\
\ttfamily ~Q~\char61{}~sum\char40{}~L~\char41{}
\end{tabbing}
\end{quote}
Each equality corresponds to a step in the path extracted from the typed LA diagram.
Type-correctness is ensured by the diagram itself.
The simplified (abstract) BNF syntax of this kernel language is
\begin{hscode}\SaveRestoreHook
\column{B}{@{}>{\hspre}l<{\hspost}@{}}%
\column{7}{@{}>{\hspre}l<{\hspost}@{}}%
\column{E}{@{}>{\hspre}l<{\hspost}@{}}%
\>[B]{}\Varid{dsl}\mathbin{:=}{(\Varid{v}\mathrel{=}\Varid{m})}^{*}{}\<[E]%
\\
\>[B]{}\Varid{m}\mathbin{:}\mathrel{=}\Varid{v}\;\asor\;\mathtt{krao}\;(\Varid{m},\Varid{m'})\;\asor{}\<[E]%
\\
\>[B]{}\hsindent{7}{}\<[7]%
\>[7]{}\mathtt{dot}\;(\Varid{m},\Varid{m'})\;\asor\;\mathtt{filter}\;(\Varid{p})\;\asor{}\<[E]%
\\
\>[B]{}\hsindent{7}{}\<[7]%
\>[7]{}\mathtt{tr}\;(\Varid{m})\;\asor\;\mathtt{lift}\;(\Varid{e})\;\asor{}\<[E]%
\\
\>[B]{}\hsindent{7}{}\<[7]%
\>[7]{}\mathtt{sum}\;(\Varid{m}){}\<[E]%
\ColumnHook
\end{hscode}\resethooks
where \ensuremath{\Varid{dsl}} is the axiom and \ensuremath{\Varid{v}}, \ensuremath{\Varid{m}}, \ensuremath{\Varid{p}} and \ensuremath{\Varid{e}} are non-terminals for
variables, matrices, attribute-level predicates and attribute level expressions, respectively.
The mapping between the combinators of this language and operations of the LA kernel
is fairly obvious.
Matrix composition \ensuremath{\Conid{M} \comp \Conid{N}} and Khatri-Rao product \ensuremath{{\Conid{M}}\kr{\Conid{N}}} are encoded by
\ensuremath{\mathtt{dot}\;(\Conid{M},\Conid{N})} and \ensuremath{\mathtt{krao}\;(\Conid{M},\Conid{N})}, respectively.
\ensuremath{\mathtt{filter}\;(\Varid{p})} represents predicate \ensuremath{\Varid{p}} in the form of a vector, recall (\ref{eq:160102a}). 
\ensuremath{\mathtt{sum}\;(\Varid{m})} implements \ensuremath{\Varid{m} \comp \conv{{\bang}}}, cf.\ (\ref{eq:180222b}).  Every cell-level
operation is lifted to the corresponding matrix operation through \ensuremath{\mathtt{lift}}.
Finally, although converse is avoided in our scripts, it is available with
syntax \ensuremath{\mathtt{tr}\;(\Varid{m})} (cf.\ transpose).

%% file: sections/6-implementation.tex
\section{Implementation}

Database environments place a strong demand for real-time query processing of always increasing volumes of data. The typed LA approach can take advantage of the powerful capabilities of current processor architectures, namely to operate on vectors and matrices. The data representation and the LA operation encoding are key factors to speedup query processing.


This section focuses on the key issues required to efficiently process and execute a query: the data representation and the key LA operations, grouped into a LA kernel. This LA kernel has been used to validate and evaluate the performance of the proposed LA approach using the TPC-H benchmark suite.

\subsection{Data representation}
    
As dimension projection matrices are very sparse, each column bearing a single non-zero value --- recall e.g.\ (\ref{eq:140403a}) ---
the chosen representation format must be memory efficient and support efficient implementations of operations on such matrices.
On the other hand, the row/column orientation is dictated by the LA operations themselves. Consider, for instance, matrix composition: 
\begin{equation}
    M \comp N =
        \meither A B
    \comp
        \msplit C D =
    A \comp C + B \comp D
    \label{eq:180223_dot}
\end{equation}


This calls for a columnar representation of matrix $M$, and a row-wise representation of $N$.
This contrasts with the Hadamard and Kronecker products, which require the same row and column-wise formats.
Concerning the Khatri-Rao product, instead of
%
\begin{equation}
    M \kr N =
        \msplit A B
    \kr
        N =
    \msplit {A \kr N} {B \kr N}
    \label{eq:180223_kr_row}
\end{equation}
one may rely on property
\begin{equation}
        \meither A B
    \kr
        \meither C D =
    \meither {A \kr C} {B \kr D}
    \label{eq:180223_kr_col}
\end{equation}
and stay with a columnar representation of both argument matrices.
This matches with the LA approach proposed in this paper, which is inherently columnar.
Thus, the adopted format for matrices is CSC, the \emph{compressed sparse column} format \cite{CSC}. It represents a sparse matrix by a set of three 1-D arrays: 
\begin{itemize}
\item
	\verb+values+, with the values of each non-zero element to be represented;
\item
	\verb+row_index+, with the row indices of each non-zero value;
\item
	\verb+column_pointer+, with size \verb|n_columns+1|, that specifies the position of the non-zero elements of a given column in the other arrays, together with the number of non-zero values of each column; the 1st value is 0 and the \verb+column_pointer[column]+ is given by adding the number of non-zero values in [column-1] to the \verb+column_pointer[column-1]+.
\end{itemize}

\vskip 1ex\noindent
Figure~\ref{fig:180223_CSC} illustrates how a matrix is compressed in CSC.

\begin{figure}[h]
\hskip -1em
\begin{tabular}{cc}
	\textbf{Sample matrix}
&
	\textbf{Same matrix in CSC format}
\\
	\ensuremath{
	\begin{bmatrix}
	{1} & {0} & \colorbox{blue!30}{3} & {0} \\
	{0}  & {0} & \colorbox{blue!30}{0}  & {7}\\
	{0}  & {0} & \colorbox{blue!30}{5} & {0} \\
	{0}  & {0} & \colorbox{blue!30}{0}  & {0}
	\end{bmatrix}
	}
&
	\begin{tabular}{rl}
	$values$ & $
	\begin{bmatrix}
	{1} & \colorbox{blue!30}{3} & \colorbox{blue!30}{5} & \phantom{|}{7}
	\end{bmatrix} $
	\\
	$row\_indices$ & $
	\begin{bmatrix}
	{0} & \colorbox{blue!30}{0} & \colorbox{blue!30}{2} & \phantom{|}{1}
	\end{bmatrix} $
	\\
	$column\_pointers$ & $
	\begin{bmatrix}
	{0} & \phantom{|}{1} & \phantom{|}\colorbox{blue!15}{1} & \colorbox{blue!30}{3} & {4}
	\end{bmatrix} $
	\end{tabular}
\end{tabular}
    \caption{A sparse matrix in CSC format}
    \label{fig:180223_CSC}
\end{figure}        

Recall that each row of the projection matrix of an attribute corresponds to a data label. The number of rows is the number of distinct labels that the attribute has.
One of the challenges of the LA approach to data representation is how to represent such \emph{data labels}, that is, how to map attribute values to unique matrix indices.

To ensure that every label maps to a distinct matrix row, they are dynamically inserted in the hash table\footnote{The Glib Hash Table is used for this purpose \cite{RAF17}.}, incrementally taking the first available integer value, in an $\sql{AUTO}$ $\sql{INCREMENT}$ fashion, as in other DBMSs. A double hashing approach is followed to improve performance, since it leads to a complexity of $\mathcal{O}(1)$ on both directions of the association.
This strategy is applied independently for each attribute. However, for referential integrity to hold among all database tables, such structures are shared by both primary and foreign key attributes.

Primary key attribute values are always mapped first. Since these keys must be $\sql{DISTINCT}$ and $\sql{NOT}$ $\sql{NULL}$, there is a \emph{bijection} between row numbers (say $\card o$) and key values (say $K$), of type \ensuremath{\card{\Varid{o}}\to \Conid{K}};
for instance, consider the $orders$ table of our running example: \small
\begin{eqnarray*}
\begin{array}{r|ccccc}&{1}&{2}&{3}&{4}&{5}
	\\\hline \hbox{2723}&0&0&0&1&0
	\\\hbox{3392}&0&0&0&0&1
	\\\hbox{4354}&0&1&0&0&0
	\\\hbox{551}&0&0&1&0&0
	\\\hbox{5699}&1&0&0&0&0
\end{array}
\end{eqnarray*}
\vskip 1ex \normalsize
By respecting the order in which key attributes arise in tables, this bijection can actually be represented by the identity matrix\footnote{Compare with a change of basis in vector space terminology.}: \small
\begin{eqnarray*}
\begin{array}{r|ccccc}&{1}&{2}&{3}&{4}&{5}
	\\\hline \hbox{5699}&1&0&0&0&0
	\\\hbox{4354}&0&1&0&0&0
	\\\hbox{551}&0&0&1&0&0
	\\\hbox{2723}&0&0&0&1&0
	\\\hbox{3392}&0&0&0&0&1
\end{array}
\end{eqnarray*} \normalsize

Therefore, by mapping primary key attribute values first, in this fashion, one can \emph{cancel out} their projection functions from LA scripts altogether, resulting in great performance gains wherever joins are evaluated.
That is, the matrix representing an equi-join is the same as the one representing the foreign key.

By contrast, measures are represented by dense vectors.



%
%
%
%
%

\subsection{Operations}

A LA kernel of the main six algebraic operators was implemented containing
three matrix products --- composition, Hadamard and Khatri-Rao ---
plus attribute filtering, matrix aggregation and the \emph{lift} operator.


Three key features lead to simplified and more efficient versions of the matrix product algorithms:
\begin{itemize}
    \item matrices are sparse: only the non-zero values need to be accounted for;
    \item if measures are introduced at a later state, such sparse matrices are Boolean (no need to access cell-values, checking for non-zeros suffices);
    \item there is at most one element per column: this avoids iterating each row index array in the CSC structure, significantly reducing memory accesses.
\end{itemize}
\vskip 1ex
These modified versions take advantage of the performance capabilities of the CSC format, leading to operations faster than those in high performance BLAS implementations.
The following paragraphs address each of these key operations.

The \emph{Hadamard product} is mainly used to combine filters, which are sparse Boolean vectors. It only needs to iterate through non-zeros of one vector and check if there is a non-zero in the same position in the other vector.

When applied to vectors, the Hadamard and \emph{Khatri-Rao} products coincide, recall
(\ref{eq:180225b}). For matrices $\rarrow n M m$ and $\rarrow p N q$, the
algorithm used in $M \times N$ (Hadamard) is the foundation of $M \kr N$
(Khatri-Rao). For each column index $i$, both algorithms iterate through $M$
(using index $j$, for instance), checking if there is an element in $N$ at row (say) $k$.
Then the new element to be inserted in the output matrix obeys the following rule:
$\cell{(jq+k)} R i = \cell j M i * \cell k N i$,
where $R=M\kr N$.





\emph{Attribute filtering} has two distinct implementations for dimensions and measures.
In dimensions, this operation has two steps: (i) apply the filter
predicate to the labels of the matrix, resulting in a sparse vector that
will be used to filter the attribute, and (ii) multiply this sparse vector
by the Boolean attribute, using matrix composition. The result of the
operation is also a sparse vector, with as many columns as the original matrix.  

Since measures are stored in dense vectors, attribute filtering here
only requires one step: to iterate over the dense vector, filtering the values
in a similar way as the first step for dimensions.

\emph{Matrix aggregation} ``folds" each row (resp.\ column) of a given matrix
over a monoidal operator --- e.g.\ a sum, a count, an average, a min or a
max operation --- by replacing the row (resp.\ column) by the aggregated value.
Thus it converts a matrix $\larrow m M n$ into a vector $\larrow 1 v n$ (resp.\ $\larrow m v 1$). 
That is, the result is a single sparse vector with as many rows (resp.\ columns) as the original number of matrix rows (resp.\ columns).

Finally, the \emph{lift} operator promotes a matrix-cell-level operation to a matrix-operation.
For instance, the Hadamard product $M \times N$ is the lifting of cell multiplication, $M + N$ is that of cell addition, and so on. It is implemented in a ``polymorphic" way, accepting operations
of different arities.

%% file: sections/7-benchmarks.tex
\section{Validation and Evaluation}
\label{sec:tpch}
%
To guide and validate the LA implementation and to compare its performance with competitor solutions, several que\-ries of the TPC-H benchmark suite were selected:
\begin{itemize}
    \item query 6: useful to validate the implementation of the most basic filter operations, namely equality, relational, and 
    between;
    \item query 3, richer than 6: also explores joins and group-by clauses;
    \item query 12 and 14: rich in filter and logical operations, such as $\sql{CASE}$, $\sql{LIKE}$, $\sql{IN}$ and $\sql{NOT}$ statements;
    \item query 11: contains sub-queries and filters after group-by ($\sql{HAVING}$);
    \item query 4: explores semi-joins ($\sql{EXISTS}$).
\end{itemize}
\vskip 1ex

Although TPC-H \emph{Composite Query-per-Hour} is the standard TPC-H performance metric, this is not adequate for a fair comparative evaluation of the LA solution. The goal of this evaluation being to assess the efficiency of the LA approach to process SQL queries on several dataset sizes and only after the datasets are loaded into RAM, execution of multiple concurrent queries is not a suitable metric here. Measuring the efficiency of disk access was deferred to a later stage.




Instead, we opted to measure, for each query, the execution times for different dataset sizes (from 1 GiB to 64 GiB) and compare the measured figures of the LA approach with those of two open-source competitor database management systems: PostgreSQL and MySQL.

%
%
%

Before presenting benchmark figures, the next subsection details the testbed environment: the measurement methodology, the server specifications and how the DBMSs were configured.


\subsection{Testbed environment}

Trustworthy experimental results must be reproducible. 
When these results are code execution times, several runs give a clue on the execution stability. 
To minimize external unwanted interference, we always look for the faster times, but more than just one value.  Our measurements are based on the 3-best runs out of 10, within a 5\% max error interval; the best of these 3 is the recorded time. 

Table 1 shows the key features of the testbed environment.





\small
\begin{table}[h!]
\caption{Testbed environment}\small
\centering
\vskip 1ex
\begin{tabular}{|l|l|} \hline
\#PU-chips & 2 \\ \hline
Model & Intel Xeon E5-2683v4 \\ \hline
Base clock freq & 2.10 GHz (up to 3.00 GHz)\\ \hline
\#PU cores & 2 x 16 (2-way SMT support)\\ \hline
L1 cache & 2 x 16 x 32 KiB\\ \hline
L2 cache & 2 x 16 x 256 KiB\\ \hline
L3 cache & 2 x 40 MiB\\ \hline
RAM & 2 x 128 GiB (NUMA)\\ \hline
OS & CentOS 6.3\\ \hline
PostgreSQL & V. 10.2\\ \hline
MySQL & V. 5.7\\ \hline
\end{tabular}
\end{table}
\normalsize


To ensure fairness, each competitor DBMS was properly configured with support from their technical staff. 

MySQL has three available storage engines: MyISAM and InnoDB are the most common ones, and MEMORY (HEAP). Since the goal in this comparative evaluation is to measure in-memory performance, the MEMORY (HEAP) alternative was chosen.

PostgreSQL has no storage engine equivalent to MEMORY (HEAP) in MySQL; a fair measure of execution times requires two runs to warm up the cache (DB-cache in RAM) before the 10 runs.
The recommended size for the shared buffers is 1/4 of the RAM size, but this value was set to 3/4 to ensure that all queries data could fit in these buffers.\footnote{Details of the overall implementation including these configurations can be found in the repository of the project: \url{https://github.com/Hubble83/LA_benchmarks}.}








\subsection{Result analysis}


This section plots and discusses measured performance: execution times and memory usage for the sequential version (single-threaded) of six selected queries from the TPC-H benchmark suite, in each of the three competing environments, PostgreSQL, MySQL and the LA approach. It also compares execution times of the sequential and parallel versions of a single query (query 6), using up to all available cores in the server. MySQL was excluded from this comparison since it has no parallel version of a single query. 

\input{plots/q3.tex}
\input{plots/q4.tex}

The plots of Figures~\ref{fig:q3} and \ref{fig:q4} show the measured times for two queries, where the scale factors have a close relationship with the dataset size (in GiB). The standard deviation is not displayed in each data-point since it is so small that stay hidden in most cases. 


The LA approach is clearly the fastest in query 3 for all dataset sizes, while the redundant filtering operations in query 4 take over 60\% of the overall execution time, considerably degrading its performance (the same happens in query 12). The removal of this bottleneck is the subject of ongoing work \cite{Af18}.

\input{plots/TIME/plot32.tex}

Since the 6 tested queries display almost the same linear behaviour across all scale factor values, there is no need to display their plots. Instead, Figure \ref{fig:time6q} compares the performance of all systems for each of the sequential versions of the six queries, for a scale factor $2^{5}$. The LA approach has proved to be faster than its competitors, in its current prototype version; it only has a lower performance in queries 4 and 12 due to redundant filtering (already explained). 

The use of column-oriented tables in the LA approach (attribute oriented) gives advantages over the competitors, namely because:
\begin{itemize}
    \item the columnar approach loads less data;
    \item it avoids operations that implicitly require a row orientation (namely, converse and matrix composition); thus the overall emphasis on the Khatri-Rao product;
    \item measures are incorporated as late as possible, taking advantage of using Boolean matrices as long as possible;
    \item it saves one matrix composition by equi-join, due to the smart encoding of primary keys.
\end{itemize}

\vskip 1ex



Performance can be further improved if the available cores in the PU-chips are adequately used: while MySQL explores parallelism by concurrently processing multiple queries, PostgreSQL can use in parallel up to all available cores to process any query, and each kernel operation in LA approach can also use up to all available cores.

\input{plots/q6par.tex}

Figure~\ref{fig:q6par} compares the single and multi-threaded versions of query 6 in the LA approach with the PostgreSQL versions. As expected, both parallel versions run consistently faster than the corresponding sequential ones, and the gain is lower when the dataset size is small. 

However, the parallel efficiency in both systems is quite low: using 32 cores, only the larger scale factor managed to reach 6x speedup. PostgreSQL is configured by default to use only 2 workers per query (equivalent to 3 threads) and the scalability analysis of the LA approach has not been performed yet. Performance tuning will follow soon.

In some queries the parallel version of PostgreSQL has unstable behaviour for larger dataset sizes; for instance, in query 3 its query planner fails in such a way that the parallel execution times in datasets larger than $2^{4}$ are longer than the sequential version. 

Overall, the LA approach shows again its parallel superiority against PostgreSQL and both versions (sequential and parallel) have a consistent behaviour through all queries.



Code efficiency is also related to the required memory to run each query. The three systems follow different approaches: while PostgreSQL loads blocks of data in RAM, keeping those that it may need (disk cache), all data in MySQL is directly inserted in RAM (in these tests). The LA approach only places in RAM the attributes it will need.

\input{plots/RAM/plot32.tex}

Figure~\ref{fig:mem6q} shows the maximum RAM space required for each of the sequential versions of the six queries in each DBMS (using the 3-worst case out of 10), for the scale factor $2^{5}$, and measured using \emph{dstat}. 

This plot clearly shows that the LA approach is very efficient in managing the used RAM and these figures can be further improved.

%% file: plots/q3.tex
\begin{figure}[ht]
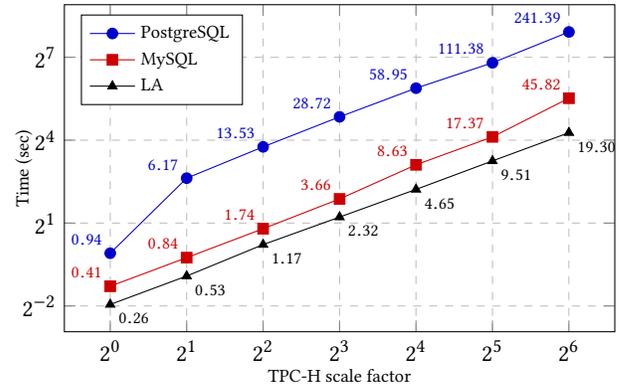

\centering
\begin{myplot}[width=1.05\linewidth, cycle list name=top_top_bot]
        \addplot table [x=dataset, y=k-best, col sep=comma] {plots/PostgreSQL/10.2/q3.csv};
        %
        \addplot table [x=dataset, y=k-best,col sep=comma] {plots/MySQL/q3.csv};
        %
        \addplot table [x=dataset, y=k-best,col sep=comma] {plots/LA/q3.csv};
        \legend{PostgreSQL,MySQL,LA};
\end{myplot}
\caption{TPC-H Query 3}
\label{fig:q3}
\end{figure}

%% file: plots/q4.tex
\begin{figure}[ht]
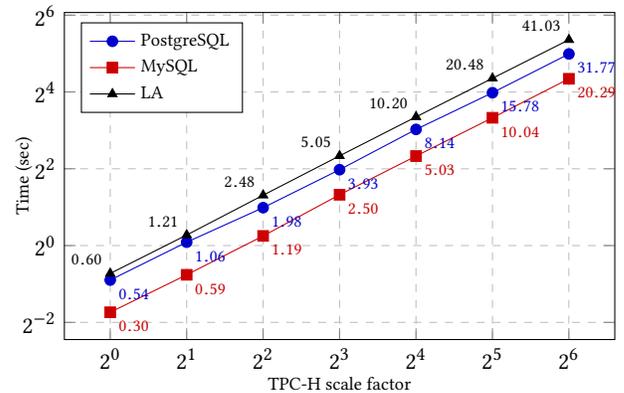

\centering
\begin{myplot}[width=1.05\linewidth, cycle list name=bot_bot_top]
        \addplot table [x=dataset, y expr=\thisrow{k-best}/1000,,col sep=comma] {plots/PostgreSQL/10.2/q4.csv};
        %
        \addplot table [x=dataset, y=k-best,col sep=comma] {plots/MySQL/q4.csv};
        %
        \addplot table [x=dataset, y=k-best,col sep=comma] {plots/LA/q4_v2.csv};
        \legend{PostgreSQL,MySQL,LA};
\end{myplot}
\caption{TPC-H Query 4}
\label{fig:q4}
\end{figure}

%% file: plots/TIME/plot32.tex
\begin{figure}[ht!]
\centering
\begin{tikzpicture}[
        hatch distance/.store in=\hatchdistance,
        hatch distance=8pt,
        hatch thickness/.store in=\hatchthickness,
        hatch thickness=2pt
    ]
    
\begin{axis}[
    ybar,
    width  = 1.05\linewidth,
    height = .6\linewidth,
    major x tick style = transparent,
    ybar=2*\pgflinewidth,
    bar width=7pt,
    ymajorgrids = true,
    ylabel = {Time (sec)},
    xlabel = {TPC-H Query},
    symbolic x coords={Q3,Q4,Q6,Q11,Q12,Q14},
    xtick = data,
    scaled y ticks = false,
    ymin=0,
    area legend,
    legend cell align=left,
    legend style={%
        at={(0.05,0.95)},%
        legend pos=north east,%
        font=\footnotesize
    },%
    enlarge x limits=0.1,
    label style={font=\footnotesize},%
    ylabel shift=-5pt,%
    xlabel shift=-3pt,
    ymax = 160,
    legend columns=3,
    ylabel shift=-5pt,%
    ymode=log,%
    log basis y=2,%
    ymin=1
]

\addplot[blue!80!black,fill=white,
postaction={pattern=north east lines,
            pattern color=blue!80!black}
            ] table [x=query, col sep=comma,
y=k-best] {plots/TIME/pg32.csv};
\addplot[red!80!black,fill=white,
postaction={pattern=horizontal lines,
            pattern color=red!80!black}] table [x=query, col sep=comma,
y=k-best] {plots/TIME/mysql32.csv};
\addplot[black,fill=white,
    postaction={pattern color=black,pattern=north west lines}
] table [x=query,col sep=comma,
y=k-best] {plots/TIME/la32.csv};

\legend{PostgreSQL,MySQL,LA}
\end{axis}



\end{tikzpicture}
\caption{Execution times (scale factor: $2^{5}$)}
\label{fig:time6q}
\end{figure}
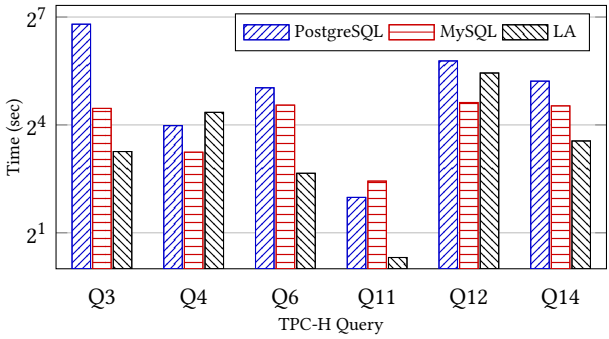

%% file: plots/q6par.tex
\begin{figure}[ht]
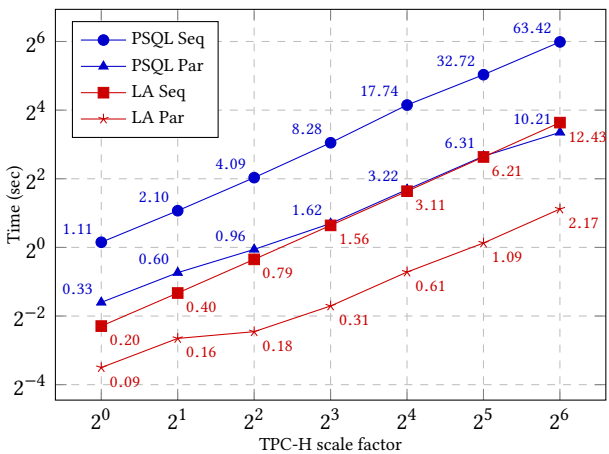

\centering
\begin{myplot}[width=1.05\linewidth, cycle list name=ttbb, height=.8\linewidth]
        \addplot table [x=dataset, y=k-best, col sep=comma] {plots/PostgreSQL/10.2/q6.csv};
        \addplot table [x=dataset, y=k-best,col sep=comma] {plots/PostgreSQL/q6par.csv};
        \addplot table [x=dataset, y=k-best,col sep=comma] {plots/LA/q6.csv};
        \addplot table [x=dataset, y=k-best,col sep=comma] {plots/LA/q6par.csv};
        \legend{PSQL Seq,PSQL Par,LA Seq, LA Par};
\end{myplot}
\caption{Sequential and Parallel Query 6}
\label{fig:q6par}
\end{figure}

%% file: plots/RAM/plot32.tex
\begin{figure}[h!]
\centering
\begin{tikzpicture}
    
\begin{axis}[
    ybar,
    width  = 1.05\linewidth,
    height = .6\linewidth,
    major x tick style = transparent,
    ybar=2*\pgflinewidth,
    bar width=7pt,
    ymajorgrids = true,
    ylabel = {Memory usage (GiB)},
    xlabel = {TPC-H Query},
    symbolic x coords={Q3,Q4,Q6,Q11,Q12,Q14},
    xtick = data,
    scaled y ticks = false,
    ymin=0,
    area legend,
    legend cell align=left,
    legend style={%
        at={(0.05,0.95)},%
        legend pos=north east,%
        font=\footnotesize
    },%
    enlarge x limits=0.1,
    label style={font=\footnotesize},%
    ylabel shift=-5pt,%
    xlabel shift=-3pt,
    ymax = 200,
    legend columns=3,
    ymode=log,%
    log basis y=2,%
    ymin=1
]

\addplot[blue!80!black,fill=white,
postaction={pattern=north east lines,
            pattern color=blue!80!black}
            ] table [x=query, col sep=comma,
y expr=\thisrow{k-best}/(1024*1024*1024)
] {plots/RAM/pg32.csv};
\addplot[red!80!black,fill=white,
postaction={pattern=horizontal lines,
            pattern color=red!80!black}] table [x=query, col sep=comma,
y expr=\thisrow{k-best}/(1024*1024*1024)
] {plots/RAM/mysql32.csv};
\addplot[black,fill=white,
    postaction={pattern color=black,pattern=north west lines}
]  table [x=query,col sep=comma,
y expr=\thisrow{k-best}/(1024*1024*1024)
] {plots/RAM/la32.csv};

\legend{PostgreSQL,MySQL,LA}
\end{axis}



\end{tikzpicture}
\caption{Memory usage (scale factor: $2^{5}$)}
\label{fig:mem6q}
\end{figure}
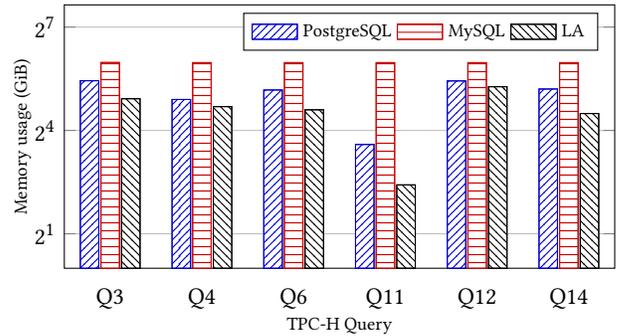

%% file: sections/8-conclusions.tex
\section{Conclusions}
This paper presents and validates a typed linear algebra (LA) approach to analytical data querying. The main novelty consists in representing data by matrices which are aggregated in a strongly typed manner, in contrast with the modest typing facilities of languages such as e.g.\ SQL, and with other LA techniques used in data analysis \cite{AM05,Ker14}, including incremental tensor analysis \cite{Sun:06,STPYF08}, which are by and large untyped.

The theory behind the overall strategy, categorial linear algebra \cite{MO13c}, enables a diagrammatic representation of both data (matrices, the arrows of diagrams) and queries (paths in diagrams involving the required dimensions and measures) in a type-safe way. The laws of LA enable a number of path transformations which preserve correctness while improving efficiency. In particular, and because the overall approach is inherently columnar, the so-called Khatri-Rao matrix product \cite{RR:98} gains prominence among the other LA operations used in queries.

A prototype of these LA kernel operations was implemented, showing our strategy performing better than other standard technologies in the majority of the tests carried out. However, there is still much room for improvement and consolidation before we can say we have a winning approach, as indicated below in the ongoing and future work section.

\subsection{Related work}
This work is a follow up of previous research in adopting typed linear algebra and its diagram-orientation to data analysis \cite{MO15,OM17}. Although 
it may seem at first sight that this approach bears some resemblance
to graph databases \cite{Wo12}, nodes in our diagrams represent data types and not individual data items.

There is much work on scalable execution models for data bases, namely in the \emph{columnar} trend. Kernert \emph{et al} \cite{Ker14,Ker15} present an approach to integrate sparse matrices in column-oriented in-memory database systems. This includes API-level support for performing elementary linear algebra operations.

Qin and Rusu \cite{Qin17} study linear algebra scalability for big model analytics. The emphasis is on gaining efficiency in dot-product joins, a primary operation for training linear models. This contrasts with the current paper, where joins are not the main problem, the focus being on the Khatr-Rao product, an useful operation often absent from the LA kernels described in the literature.

Finally note that the projection matrices used in this research also intersect with the bitmaps used to represent table indexes. Wu \emph{et al} \cite{WOS06} propose a way to overtake the large memory demands of the IBM's Model 204 bitmap representation \cite{ON89}. However, our sparse representation inherently compresses the data, and so such efforts are not required.






\subsection{Ongoing and future work}
\begin{figure}[ht]
    \centering
	\includegraphics[width=\columnwidth]{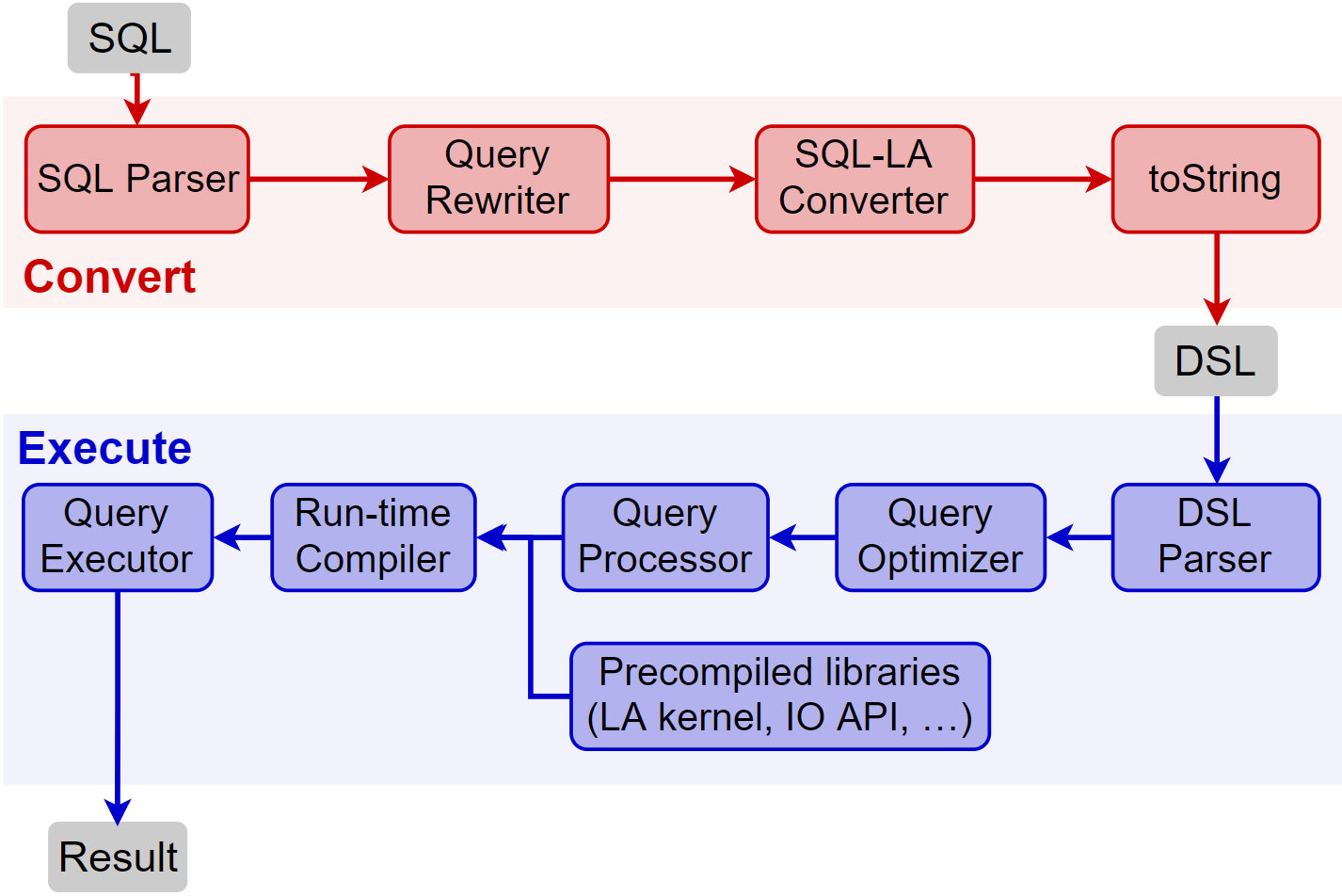}
    \caption{The framework architecture}
    \label{fig:frame}
\end{figure}

This paper validates and evaluates a proof of concept framework for analytical querying on a strong formal basis, with efficiency gains. 
However, to convert this proof of concept into an usable software toolset, additional effort is required. The implementation of the framework architecture shown in Figure \ref{fig:frame} is on-going work \cite{AF17,Af18}. 





Moreover, there is still room to improve the parallel efficiency of the overall design and implementation:
\begin{itemize}
    \item to evaluate the scalability of the LA operations to define the max number of sustainable concurrent cores per query without degrading performance;
    \item to explore matrix splitting into blocks of columns to allow operations in micro batching or even streaming \cite{akidau2015dataflow}, if the blocks are short enough; this not only will speedup most LA operations with vectors and matrices, but also require less memory usage;
    \item to dynamically explore parallelism in the execution of independent pipeline stages for any heterogeneous server architecture, with or without computing accelerators (e.g., GPU, FPGA) \cite{carbone2015apache,Pereira2016}; 
    \item to implement the LA kernels in distributed memory; by splitting or replicating matrices across servers, attributes larger than single server memories can be que\-ried.
\end{itemize}
\vskip 1ex
In another direction, we would like to study the practical impact of incrementality (section \ref{sec:180301a}) on efficient data cube updating, combining the results of this paper with those of reference \cite{OM17}.

%% file: sections/appendix.tex

\newenvironment{nappendix}{\appendix\section*{Appendix}}{}

\begin{nappendix}
Two appendices are included: one describing the typed linear algebra pontwise notation adopted in the paper and the other about the so-called "pairing wheel" rule, which can be used to improve the performance of queries involving the Khatri-Rao product.

\section{Typed linear algebra pointwise notation} \label{sec:180220a}

The notation adopted for expressing index-wise matrix expressions is known as
Eindhoven quantifier calculus, see e.g.\ \cite{BM06}.
In this notation, a quantified expression is always of the form
\begin{eqnarray*}
	\ensuremath{\rcb{{\sum}}{\Varid{x}}{\Conid{R}}{T }}
\end{eqnarray*}
where \ensuremath{\Conid{R}} is a predicate specifying the \emph{range} of the quantification and \ensuremath{T } is a numeric term.
\
In case \ensuremath{T \mathrel{=}\Conid{B} \times \Conid{M}} where Boolean \ensuremath{\Conid{B}\mathrel{=}\mathopen\langle \Conid{P}\mathclose\rangle } encodes predicate \ensuremath{\Conid{P}}
--- recall (\ref{eq:160130a}) ---
we have the \emph{trading rule}:
\begin{eqnarray}
	\ensuremath{\rcb{{\sum}}{\Varid{x}}{\Conid{R}}{\mathopen\langle \Conid{P}\mathclose\rangle  \times \Conid{M}}} \wider= \ensuremath{\rcb{{\sum}}{\Varid{x}}{\Conid{R}\mathrel{\wedge}\Conid{P}}{\Conid{M}}}
	\label{eq:151229a}
\end{eqnarray}
Another (very) useful law of this calculus is known as the \emph{one-point} rule
\begin{eqnarray}
	\ensuremath{\rcb{{\sum}}{\Varid{k}}{\Varid{k}\mathrel{=}\Varid{e}}{T }} &=& T[k := e]
	\label{eq:130228b-LA}
\end{eqnarray}
where expression $T[k := e]$ denotes \ensuremath{T } with every occurrence of \ensuremath{\Varid{k}} replaced by \ensuremath{\Varid{e}}.

The LA (index-free) properties given in this paper, namely (\ref{eq:120428d},
 \ref{eq:120428c}, \ref{eq:140321b}) can be derived from (\ref{eq:151229a})
and (\ref{eq:130228b-LA}).

\section{The \aspas{pairing wheel} rule}  \label{sec:18022b}
In typed LA, the information captured by the three matrices \ensuremath{\Conid{M}}, \ensuremath{\Conid{P}} and \ensuremath{\Conid{Q}} in
\begin{eqnarray}
\myxym{
&
	\ensuremath{\Conid{B}}
\\
&
	\ensuremath{\Conid{A}}
		\ar[u]_{M}
		\ar[dr]^{P}
		\ar[dl]_{Q}
\\
	\ensuremath{\Conid{C}}
&&
	\ensuremath{\Conid{D}}
}
	\label{eq:180228e}
\end{eqnarray}
can be aggregated in several ways, namely
\begin{quote}
	\ensuremath{\longrarrow{\Conid{B}}{({\Conid{P}}\kr{\Conid{Q}}) \comp \conv{\Conid{M}}}{\Conid{D} \times \Conid{C}}}
\\
	\ensuremath{\longrarrow{\Conid{D}}{({\Conid{Q}}\kr{\Conid{M}}) \comp \conv{\Conid{P}}}{\Conid{C} \times \Conid{B}}}
\\
	\ensuremath{\longrarrow{\Conid{C}}{({\Conid{M}}\kr{\Conid{P}}) \comp \conv{\Conid{Q}}}{\Conid{B} \times \Conid{C}}}
\end{quote}
all isomorphic to each other:
\begin{eqnarray*}
\xymatrix@R=3em@C=-2em{
&
	\ensuremath{\Conid{B}\to \Conid{D} \times \Conid{C}}
        	\ar@/^1pc/[dr]^-{\ensuremath{\alpha }}
\\
	\ensuremath{\Conid{C}\to \Conid{B} \times \Conid{D}}
        	\ar@/^1pc/[ur]^-{\ensuremath{\alpha }}
&&
	\ensuremath{\Conid{D}\to \Conid{C} \times \Conid{B}}
        	\ar@/^2pc/[ll]^-{\ensuremath{\alpha }}
}
\end{eqnarray*}
The rotation among matrices and types justifies the name ``pairing wheel" given to
(\ref{eq:180228e}). 
Isomorphism \ensuremath{\alpha } holds in the sense that every cell of one of the aggregates is uniquely
represented by another cell in any other aggregate, for instance:
\begin{eqnarray*}
\start
	\ensuremath{(\Varid{d},\Varid{c})\;(({\Conid{P}}\kr{\Conid{Q}}) \comp \conv{\Conid{M}})\;\Varid{b}}
\just={ composition ; Khatri-Rao}
	\ensuremath{\rcb{{\sum}}{\Varid{a}}{}{(\cell{\Varid{d}}{\Conid{P}}{\Varid{a}} \times \cell{\Varid{c}}{\Conid{Q}}{\Varid{a}}) \times \cell{\Varid{a}}{\conv{\Conid{M}}}{\Varid{b}}}}
\just={ converse; \ensuremath{ \times } is associative and commutative}
	\ensuremath{\rcb{{\sum}}{\Varid{a}}{}{(\cell{\Varid{c}}{\Conid{Q}}{\Varid{a}} \times \cell{\Varid{b}}{\Conid{M}}{\Varid{a}}) \times \cell{\Varid{a}}{\conv{\Conid{P}}}{\Varid{d}}}}
\just={ composition ; Khatri-Rao}
	\ensuremath{(\Varid{c},\Varid{b})\;(({\Conid{Q}}\kr{\Conid{M}}) \comp \conv{\Conid{P}})\;\Varid{d}}
\end{eqnarray*}
Thus: \ensuremath{\alpha \;(({\Conid{P}}\kr{\Conid{Q}}) \comp \conv{\Conid{M}})\mathrel{=}({\Conid{Q}}\kr{\Conid{M}}) \comp \conv{\Conid{P}}}. In the special case of one of the
matrices being a vector, say \ensuremath{\Conid{M}\mathrel{=}\Varid{v}} in
\begin{eqnarray*}
\xymatrix{
&
	\ensuremath{\mathrm{1}}
\\
&
	\ensuremath{\Conid{A}}
		\ar[u]_{v}
		\ar[dr]^{P}
		\ar[dl]_{Q}
\\
	\ensuremath{\Conid{C}}
&&
	\ensuremath{\Conid{D}}
}
\end{eqnarray*}
we get that
\begin{eqnarray}
	\ensuremath{\longrarrow{\mathrm{1}}{({\Conid{P}}\kr{\Conid{Q}}) \comp \conv{\Varid{v}}}{\Conid{D} \times \Conid{C}}}
	\label{eq:180222d}
\end{eqnarray}
bears the same information as
\begin{eqnarray}
	\ensuremath{\longlarrow{\Conid{D}}{({\Conid{Q}}\kr{\Varid{v}}) \comp \conv{\Conid{P}}}{\Conid{C}}}
	\label{eq:180222e}
\end{eqnarray}
which is the converse of
\begin{eqnarray*}
	\ensuremath{\longrarrow{\Conid{C}}{({\Varid{v}}\kr{\Conid{P}}) \comp \conv{\Conid{Q}}}{\Conid{D}}}
\end{eqnarray*}
Thus the rule:
\begin{eqnarray}
	\ensuremath{\Conid{P} \comp \conv{({\Conid{Q}}\kr{\Varid{v}})}\mathrel{=}({\Varid{v}}\kr{\Conid{P}}) \comp \conv{\Conid{Q}}}
	\label{eq:180222c}
\end{eqnarray}
For \ensuremath{\Conid{Q}\mathrel{=}{\bang}} in (\ref{eq:180222c}) we have, via (\ref{eq:140602a}),
\begin{eqnarray}
	\ensuremath{\Conid{P} \comp \conv{\Varid{v}}\mathrel{=}({\Varid{v}}\kr{\Conid{P}}) \comp \conv{{\bang}}}
	\label{eq:180222b}
\end{eqnarray}
Moreover, for \ensuremath{\Conid{Q}\mathrel{=}\Varid{id}} in (\ref{eq:180222c}), we get
\begin{eqnarray*}
	\ensuremath{\Conid{P} \comp \conv{({\Varid{id}}\kr{\Varid{v}})}\mathrel{=}{\Varid{v}}\kr{\Conid{P}}}
\end{eqnarray*}
Finally, since \ensuremath{{\Varid{id}}\kr{\Varid{v}}} is diagonal and therefore symmetric,
\begin{eqnarray}
	\ensuremath{\Conid{P} \comp ({\Varid{id}}\kr{\Varid{v}})\mathrel{=}{\Varid{v}}\kr{\Conid{P}}}
	\label{eq:160101a}
\end{eqnarray}
and, of course, for \ensuremath{\Conid{P}\mathrel{=}\Varid{id}},
\begin{eqnarray}
	\ensuremath{{\Varid{id}}\kr{\Varid{v}}\mathrel{=}{\Varid{v}}\kr{\Varid{id}}}
\end{eqnarray}

\end{nappendix}